\documentclass[conference,compsoc]{IEEEtran}
\newcommand{\descr}[1]{\smallskip \noindent \textbf{#1}}

\usepackage[algo2e,ruled]{algorithm2e}
\usepackage{color}
\usepackage{hyperref}
\usepackage[moderate,tracking=normal]{savetrees}
\usepackage{xspace}
\usepackage{enumerate}
\usepackage{booktabs}
\usepackage[export]{adjustbox}
\usepackage{multicol}
\usepackage[table,xcdraw]{xcolor}
\usepackage[shortlabels]{enumitem}
\usepackage{amssymb}
\usepackage{amsmath}
\usepackage{mathtools}
\usepackage{soul}
\DeclarePairedDelimiter\ceil{\lceil}{\rceil}

\allowdisplaybreaks 
\usepackage{amsthm} 
\usepackage{graphicx}
\usepackage[hang,flushmargin]{footmisc}
\usepackage{lipsum}
\usepackage{comment}
\usepackage{xcolor}
\usepackage{alltt}
\usepackage{algorithm}
\usepackage{algorithmic}
\usepackage{setspace}
\usepackage[caption=false,font=footnotesize]{subfig}
\makeatletter
\def\hlinewd#1{%
  \noalign{\ifnum0=`}\fi\hrule \@height #1 \futurelet
   \reserved@a\@xhline}
\makeatother
\usepackage{tabularx}
\usepackage[T1]{fontenc}
\usepackage[utf8]{inputenc}
\usepackage{bm}
\usepackage{multirow}
\usepackage{newfloat}
\usepackage[normalem]{ulem}
\DeclareCaptionType{copyrightbox}

\newtheorem{proposition}{Proposition}

\usepackage{pifont}
\newcommand{\cmark}{\ding{51}}%
\newcommand{\xmark}{\ding{55}}%
\newcommand{\Hquad}{\hspace{0.5em}}

\makeatletter
\newcounter{protocol}
\newenvironment{protocol}[1]
{\par\addvspace{\topsep}
	\noindent
	\tabularx{\linewidth}{@{} X @{}}
	\hline \vspace{-0.5em}
	\refstepcounter{protocol}\textbf{Protocol \theprotocol} #1 \\
	\hline
}
{ \\
	\hline
	\endtabularx
	\par\addvspace{\topsep}}

\makeatletter
\newcounter{method}
\newenvironment{method}[1]
{\par\addvspace{\topsep}
	\noindent
	\tabularx{\linewidth}{@{} X @{}}
	\hline \vspace{-0.5em}
	\refstepcounter{method}\textbf{M\themethod}#1 \\
	\hline
}
{ \\
	\hline
	\endtabularx
	\par\addvspace{\topsep}}

\newcommand{\sbline}{\\[.5\normalbaselineskip]}

\newcommand{\sys}{\textsc{sf-pca}\xspace}

\newif\ifcomment
\commentfalse

\definecolor{cerise}{rgb}{0.87, 0.19, 0.39}
\definecolor{cadmiumgreen}{rgb}{0.0, 0.42, 0.24}

\ifcomment
	\newcommand{\ap}[1]{\textbf{\em\color{blue}#1}}
	\newcommand{\df}[1]{{\color{blue}#1}}
	\newcommand{\strat}[1]{{\color{orange}#1}}
	\newcommand{\rem}[1]{{\textcolor{red}{\sout{#1}}}}
	\newcommand{\remtit}[1]{{\color{blue}#1}}

	\newcommand{\spa}[1]{{\color{cadmiumgreen}#1}}
	
	\newcommand{\jpb}[1]{\textbf{\em\color{cerise}[JPB: #1]}}
\else  
    \newcommand\ap[1]{}
    \newcommand\df[1]{#1}
    \newcommand\rem[1]{}
    \newcommand\remtit[1]{}
    \newcommand\strat[1]{}
	
    \newcommand\sinem[1]{}
    \newcommand\spa[1]{}
    
    \newcommand\jpb[1]{}
\fi
\usepackage[font=footnotesize]{caption}

\begin{document}
%
\title{Scalable and Privacy-Preserving Federated Principal Component Analysis}

\makeatletter
\newcommand{\linebreakand}{%
  \end{@IEEEauthorhalign}
  \hfill\mbox{}\par
  \mbox{}\begin{@IEEEauthorhalign}
}
\makeatother

\author{
\IEEEauthorblockN{David Froelicher\IEEEauthorrefmark{1}\,\IEEEauthorrefmark{2}\,\IEEEauthorrefmark{3},
Hyunghoon Cho\IEEEauthorrefmark{1}\,\IEEEauthorrefmark{3},
Manaswitha Edupalli\IEEEauthorrefmark{3},
Joao Sa Sousa\IEEEauthorrefmark{4},
Jean-Philippe Bossuat\IEEEauthorrefmark{5},}
\IEEEauthorblockN{
Apostolos Pyrgelis\IEEEauthorrefmark{4},
Juan R. Troncoso-Pastoriza\IEEEauthorrefmark{5},
Bonnie Berger\IEEEauthorrefmark{2} and
Jean-Pierre Hubaux\IEEEauthorrefmark{4}\,\IEEEauthorrefmark{5}}
\IEEEauthorblockA{
\IEEEauthorrefmark{2}MIT, \IEEEauthorrefmark{3}Broad Institute of MIT and Harvard, \IEEEauthorrefmark{4}EPFL, \IEEEauthorrefmark{5}Tune Insight SA}   
}



\maketitle

\begin{abstract}
Principal component analysis (PCA) is an essential algorithm for dimensionality reduction in many data science domains.
We address the problem of performing a federated PCA on private data distributed among multiple data providers while ensuring data confidentiality. Our solution, \sys, is an end-to-end secure system that 
preserves the confidentiality of both the original data and all intermediate results in a passive-adversary model with up to all-but-one colluding parties.
\sys jointly leverages multiparty homomorphic encryption, interactive protocols, and edge computing to efficiently interleave computations on local cleartext data with operations on collectively encrypted data.
\sys obtains results as accurate as non-secure centralized solutions, independently of the data distribution among the parties. It scales linearly or better with the dataset dimensions and with the number of data providers.
\sys is more precise than existing approaches that approximate the solution by combining local analysis results, and between 3x and 250x faster than privacy-preserving alternatives based solely on secure multiparty computation or homomorphic encryption. 
Our work demonstrates the practical applicability of secure and federated PCA on private distributed datasets.
\end{abstract}

\setcounter{page}{1}
\thispagestyle{plain}



\section{Introduction}\label{intro}

\let\thefootnote\relax\footnotetext{* Equal contribution}
Principal component analysis (PCA) \cite{hotelling1933analysis,pearson1901liii} is an algorithm for analyzing a high-dimensional dataset, represented as a matrix of samples (rows) by features (columns), to uncover a small set of orthogonal directions---principal components (PCs)---that together maximally capture the observed variance among the data samples. Given the ability of PCA to reduce the dimensionality of a dataset while preserving its information content, it is commonly used in many data analysis workflows, including predictive modeling and exploratory data analysis (e.g., clustering and data visualization)~\cite{applications1,giri2013automated,gumus2010evaluation,jolliffe2016principal, martis2013ecg, pasini2017principal,price2006principal,vozalis2007recommender}. PCA is also a common pre-processing technique in machine learning (ML) pipelines, where the goal is to reduce the number of features to avoid overfitting and improve generalization performance \cite{martis2013ecg, accuracy2, accuracy3, accuracy4, jing2015svm}. While more sophisticated non-linear dimension-reduction approaches have been proposed (\df{e.g., }based on autoencoders \cite{hinton2006reducing, roweis2000nonlinear}), PCA remains the de-facto standard method for dimension reduction, as it is computationally efficient, theoretically well-understood, and reliably accurate~\cite{jolliffe2016principal, fodor2002survey}.

Many modern applications of PCA involve data from individuals, raising privacy-related challenges that limit the availability of data for such analyses.
In the biomedical domain, the high-dimensional nature of biomedical measurements often necessitate the use of PCA to extract key features from personal data, including genetic sequences~\cite{price2006principal,freedman2004assessing}, single-cell transcriptomic data~\cite{hie2020computational,sav2022privacy}, medical images 
\cite{giri2013automated,bouwmans2018applications} 
 and time-series data 
~\cite{martis2013ecg,wilaiprasitporn2019affective}. 
PCA is also commonly used in other domains involving personal data, including quantitative finance~\cite{pasini2017principal} and recommender systems~\cite{vozalis2007recommender}.
Due to the privacy and security implications, the sharing of personal data in these domains is often prohibited, rendering the data analysis difficult or even impossible. This results in sensitive data remaining siloed in access-controlled repositories and not shared across organizations, which often hinders research, innovation, and routine organizational tasks \cite{theverge}.

Federated privacy-preserving analytics, which aims to facilitate the joint analysis of sensitive data held by multiple parties using privacy-enhancing technologies~\cite{spindle,mohassel2017secureml,sav2020poseidon,zheng2019helen, froelicher2021truly}, has emerged as a promising solution to the aforementioned challenges with the potential to overcome regulatory barriers in data sharing \cite{scheibner2020revolutionizing}. 
Despite \df{the growing} interest, many essential tools for data analysis including the PCA, especially those upstream of widely studied tasks such as model training and inference, have received limited interests and are often omitted from \df{federated} workflows.
This creates an important gap in secure analytics, potentially undermining their security or utility if one falls back on a non-secure or less-accurate alternative in order to perform the full analysis.

A key challenge in developing a secure federated solution for PCA is that it requires complex and iterative computations (e.g. matrix factorization), which are costly given a large-scale input. These operations are not directly amenable to efficient computation with generic cryptographic techniques \cite{keller2018overdrive, bogdanov2008sharemind,hastings2019sok}. Reflecting this difficulty, many existing federated solutions \cite{abu2002distributed, bai2005principal, Balcan2014,Cheung, fan2019distributed, fellus2014dimensionality,gang2019fast,liang2013distributed, liang2014improved,qi2004global,Won2016}, propose that the data providers (DPs) independently perform an initial dimension reduction on their local data, before they combine their intermediate results and execute the final decomposition on the merged results. This approach, which we refer to as \textit{meta-analysis}, results in a loss of accuracy as it alters the original PCA problem and is prone to overlooking patterns spanning multiple DPs' datasets, especially when the data distributions differ among the DPs. Furthermore, most meta-analysis solutions require the DPs' intermediate results to be revealed to an aggregator server (or to other DPs) hence are not end-to-end secure. Other existing PCA solutions based on secure multiparty computation (SMC) techniques~\cite{Won2016, Bogdanov2018, cho2018secure, fan2021ppca} require the entire input data to be securely shared with a few computing servers.
With the high communication overhead of SMC, these solutions have difficulty supporting a large number of parties.

In this paper, we propose an efficient and secure system for performing a federated PCA on a distributed dataset, where the data remains protected and locally stored by the respective DPs. Our solution, named \sys (for Secure Federated PCA), executes the randomized PCA (RPCA) algorithm~\cite{halko2011finding}, the de facto standard for PCA on large-scale matrices, in a federated manner using a multiparty extension of homomorphic encryption~\cite{mouchet2019distributedbfv}.
Contrary to meta-analysis solutions, \sys directly executes a standard PCA algorithm (i.e., RPCA) to achieve state-of-the-art accuracy similar to a centralized analysis, while ensuring end-to-end privacy by protecting even the intermediate results.
Unlike SMC solutions, \sys is more communication-efficient and can be used by a large number of DPs.
\df{Note that our setting is related to cross-silo federated learning \cite{kairouz2021advances}, except we do not focus on predictive model training and we use cryptographic techniques to provide end-to-end privacy.}

Specifically, \sys is built upon the cryptographic framework of multiparty homomorphic encryption (MHE; see \S.\ref{subsec:mhe}).
In MHE, analogous to related works on threshold HE~\cite{asharov2012multiparty,cramer2001multiparty,lopez2011cloud,mouchet2022efficient}, the collective secret (or decryption) key is secret-shared among all the DPs, and the corresponding public key and additional evaluation keys required for homomorphic operations are known by all DPs. 
This ensures that, while encryption and ciphertext computations can be independently performed by each DP, decrypting ciphertexts requires all DPs to collaborate~\cite{mouchet2019distributedbfv}.
MHE's ability to offload certain computations to be locally performed by each party using the cleartext data leads to key performance improvements, as we show in our work. 
Performing a compute-intensive algorithm like RPCA, which involves sophisticated linear algebra operations (e.g., orthogonalization and eigendecomposition) \df{on input vectors and matrices of a wide range of dimensions
}, while efficiently working within the constraints of MHE and maximally exploiting its strengths is the key challenge we address in \sys by introducing optimization strategies and efficient MHE linear algebra routines.

Our evaluation demonstrates the practical performance of \sys on six real datasets.
For example, \sys securely computes five PCs on the MNIST dataset \cite{MNIST} with 60,000 samples and 760 features, split among six DPs, in 2.22 hours. In the same setting, it obtains the two PCs from a lung cancer dataset \cite{lungcancer} with 9,098 patients and 23,724 genomic features in 3.5 hours.
\sys scales at most linearly with the input dimensions and with the number of DPs. \sys is one to two orders of magnitude faster than a centralized-HE solution. It is up to ten times faster than existing SMC solutions \cite{cho2018secure}, which scale poorly with the number of DPs. We also show that \sys is highly accurate, resulting in Pearson correlation coefficients of above 0.9 (compared to the ground truth) in all settings, whereas meta-analysis often obtains inaccurate results (e.g., a correlation below 0.75 for both datasets mentioned above).
Moreover, \sys executes PCA while ensuring end-to-end data confidentiality as long as one DP is honest, whereas meta-analysis reveals the intermediate results to the aggregator server.
\df{Both centralized-HE and the previous SMC solution \cite{cho2018secure} require an honest third-party to hold the decryption key or to distribute correlated randomness for efficiency, respectively}. 

In this work, we make the following contributions:
\begin{itemize}[leftmargin=*]
    \setlength\itemsep{-0.0em}
    \item We propose \sys, a system for an efficient, federated, and end-to-end confidential execution of PCA~\cite{halko2011finding}.
    \item We demonstrate key design strategies underlying the practical performance of \sys, including: (i) maximizing operations on the DPs' cleartext local data by restructuring the computation and (ii) developing efficient linear algebra routines under a consistent vectorized encoding scheme for encrypted matrices to fully utilize the packing and \textit{single-instruction multiple-data} (SIMD) property of MHE without costly encoding conversion.
    \item \df{We introduce an adaptive approach for choosing both the high-level computational approach for PCA and the low-level MHE routines to maximize efficiency, based on the input dimensions for each computational step.} \item We propose efficient MHE-based algorithms for sophisticated linear algebra operations on encrypted matrices, including matrix multiplication, factorization, and orthogonalization, in the federated setting.
    \item We demonstrate the practical performance of \sys on six real datasets and illustrate its utility for biomedical data analysis. We show that \sys is more scalable than existing solutions for privacy-preserving PCA while producing accurate results comparable to a centralized execution of PCA regardless of the data distribution among the parties.
\end{itemize}
To the best of our knowledge, \sys is the first system to enable federated PCA in a scalable and end-to-end confidential manner. 
We note that \sys's optimization strategies and linear algebra building blocks are broadly applicable to the development of secure federated algorithms and thus are of independent interest.
\section{Related Work}\label{sec:related}
 

\subsection{Homomorphic Encryption (HE)}
We discuss prior works on linear algebra in HE and on distributed HE schemes, two essential components of \sys (\S.\ref{subsec:mheops}).

\descr{HE for Linear Algebra.}
Multiple works have shown how to optimize matrix-vector multiplications \cite{spindle,gazelle} and multiplications between small matrices (i.e., fitting in a single ciphertext) \cite{halevi2014algorithms,HaleviHelibImprove,jiang2018secure,Pradeep2021}. Multiplication of large encrypted matrices, whose rows do not fit into single ciphertexts, has been less studied. PCA requires multiple types of multiplications involving large matrices of varying dimensions, and efficiently performing these operations under encryption is key to achieving practical performance. \sys jointly leverages a range of matrix multiplication methods whose complexities scale differently with the input dimensions, making an adaptive choice for each computational step in RPCA (\S.\ref{subsub:matmult}). 


\descr{Distributed HE.}
When multiple parties use HE to combine their private data, they can either share all of their data encrypted under the same key held by a trusted entity (e.g., in a centralized scheme \cite{cheon2017homomorphic, fan2012somewhat}), or adopt a distributed scheme where no single entity holds the decryption key. In threshold encryption schemes \cite{desmedt1994threshold, zhu2020practical}, the encryption key is known to all parties whereas the decryption key is secret-shared among the parties such that a predefined number of them must collaborate to decrypt a ciphertext. In multi-key \cite{kim2022asymptotically} schemes (including a hybrid with threshold schemes \cite{kwak2021unified}), the parties have their own key pair and can jointly compute on data encrypted under different keys, but the complexity scales with the number of parties. In \sys, we rely on a multiparty HE scheme (MHE) proposed by Mouchet et al. \cite{mouchet2019distributedbfv}, which corresponds to an \textit{$s$-out-of-$s$} threshold scheme. This scheme enables local computation with complexity independent of the number of parties and provides a lightweight, interactive protocol to refresh (bootstrap) a ciphertext---a key factor for \sys's efficiency in contrast to alternative approaches (see \S.\ref{sec:design}).
\subsection{\df{Principal Component Analysis (PCA)}}
\descr{Secure Centralized PCA.}
Few solutions have been proposed for the secure centralized computation of PCA due to its computational complexity. Pereiral and Aranhal \cite{pereira2016principal} proposed a method for performing PCA on an encrypted dataset using homomorphic encryption (HE). HE-based solutions typically incur a high computational overhead compared to their cleartext counterparts. In addition, they require a costly centralization of the data and have a single point-of-failure, i.e., the holder of the decryption key. 
In \sys, since the exchanged data are encrypted with a collective key, no single entity can decrypt them, and compute-intensive HE operations (e.g., bootstrapping) are replaced by lightweight interactive protocols. \df{In \S.\ref{subsec:comparison}, we compare \sys with an HE-based centralized solution.}

\descr{Non-Secure Federated PCA.}
Solutions that enable PCA on distributed data without privacy protection fall in two main categories: iterative \cite{de2018accelerated,garber2017communication,hartebrodt2022federated,wu2018review} and non-iterative \cite{abu2002distributed,bai2005principal,Balcan2014,Cheung,fan2019distributed,fellus2014dimensionality, gang2019fast, liang2013distributed,liang2014improved, qi2004global, Won2016}. In the former, the DPs communicate and collaborate in order to perform each step of the algorithm. In the latter, the DPs perform the decomposition locally and then merge their results; we also refer to this approach as \emph{meta-analysis}. Meta-analysis requires less communication but introduces inaccuracies by approximating PCA with two levels of decomposition, i.e., an independent local decomposition by each DP and a global one for the merged results. 
These solutions typically require that the local data distribution be consistent across DPs to obtain accurate results. In addition, they are not end-to-end secure as they require the DPs' intermediate results to be revealed to an aggregator server (or to other DPs), representing a single point of failure. Intermediate results have been shown to reveal information about the original data in federated settings, e.g., in PCA \cite{hartebrodt2022federated} and ML \cite{nasr2019comprehensive,melis2019exploiting}.
\df{In contrast, \sys implicitly performs RPCA on the joint data without altering the original approach, thus obtaining accurate results independently of the data distribution among the DPs (\S.\ref{sec:eval}).} It also keeps all the exchanged information secret and does not rely on an aggregator server.

\descr{SMC-based PCA.}
Several solutions \cite{Won2016, Bogdanov2018, cho2018secure, fan2021ppca} leverage secure multiparty computation (SMC) to perform PCA on data that are secret-shared among a limited number of parties (e.g., three). These solutions require the data to be outsourced to computing parties, incurring a high communication overhead for large datasets.
Unlike SMC solutions, \sys can be efficiently used by a large number of parties, and their data are kept locally with a minimal amount of encrypted information exchanged for the PCA computation.
In \S.\ref{sec:extensions}, we discuss an extension of \sys where SMC techniques are integrated into our system to aid in carrying out non-polynomial function evaluations on small-dimensional inputs.

\descr{HE-based PCA.}
To our knowledge, Liu et al.~\cite{Liu2020} proposed the only existing homomorphic encryption (HE)-based solution for federated PCA.
However, they rely on an aggregator server that decrypts the aggregated values at each step of the process.
Since the intermediate results can reveal information about the parties' local data, these methods are not end-to-end secure.
\sys demonstrates that a fully decentralized and end-to-end secure solution for PCA is practically feasible.

\descr{Differential Privacy-based PCA.}
\df{Solutions based on differential privacy \cite{grammenos2019federated, Imtiaz2018, Wang2020} fundamentally differ from \sys in that their goal is to limit the privacy leakage of the intermediate or final results. To achieve this goal, these solutions introduce noise into the computation, making the final results less accurate.} \df{Furthermore, analogous to meta-analysis, some of these solutions rely on a local decomposition followed by a global aggregation of results, introducing an approximation error in addition to the noise added for differential privacy.}
In \sys, no intermediate result is revealed, hence differential privacy is not needed to protect the information exchanged during the algorithm.
On the other hand, if the DPs wish to reveal the final PCA result with differential privacy, such guarantee can be added to \sys \df{(\S.\ref{sec:extensions})}.
\section{Background}\label{sec:back}

\noindent \textbf{Notation.}
Matrices and vectors are denoted by boldface uppercase and lowercase characters, respectively. The $i$-th row (resp. column) of a matrix $\bm{X}^{(a\times b)}$ with $a$ rows and $b$ columns is denoted by $\bm{X}[i,:]$ (resp. $\bm{X}[:,i]$). The submatrix from (resp. up to but not including) row $i$ and column $j$ is denoted as $\bm{X}[i$:, $j$:$]$ (resp. $\bm{X}[$:$i$,:$j]$). The $i$-th element of a vector of $b$ elements $\bm{y}^{(b \times 1)}$ is denoted by $\bm{y}[i]$. Cleartext data are indicated by a tilde (e.g., $\bm{\tilde{X}}$). A matrix multiplication is denoted by $\times$.

\noindent \textbf{Principal Component Analysis (PCA).}
\label{subsec_randoPCA}
PCA is used to extract the most prominent set of linearly independent directions, i.e., principal components (PCs), that underlie a set of correlated features (columns of a data matrix). The PCs are identified in a descending order of the variance among the data points that each one captures. The PCs can be viewed as the leading eigenvectors of the feature covariance matrix, where the corresponding eigenvalues represent the variance explained. Dimension reduction of the dataset can be achieved by projecting the data points onto the PCs. Formally, PCA takes the matrix $\bm{\tilde{A}}^{(n \times m)}$ and outputs the reduced matrix $\bm{\tilde{A}'}^{(n \times \psi)}$ obtained from the projection of the input matrix onto its $\psi$ (with $\psi \ll m$) PCs.

\begin{figure}[ht]
\small
\ifcomment
 
 \else
 \vspace{-0.5em}
 \fi
    \centering
    \includegraphics[width=1.0\columnwidth]{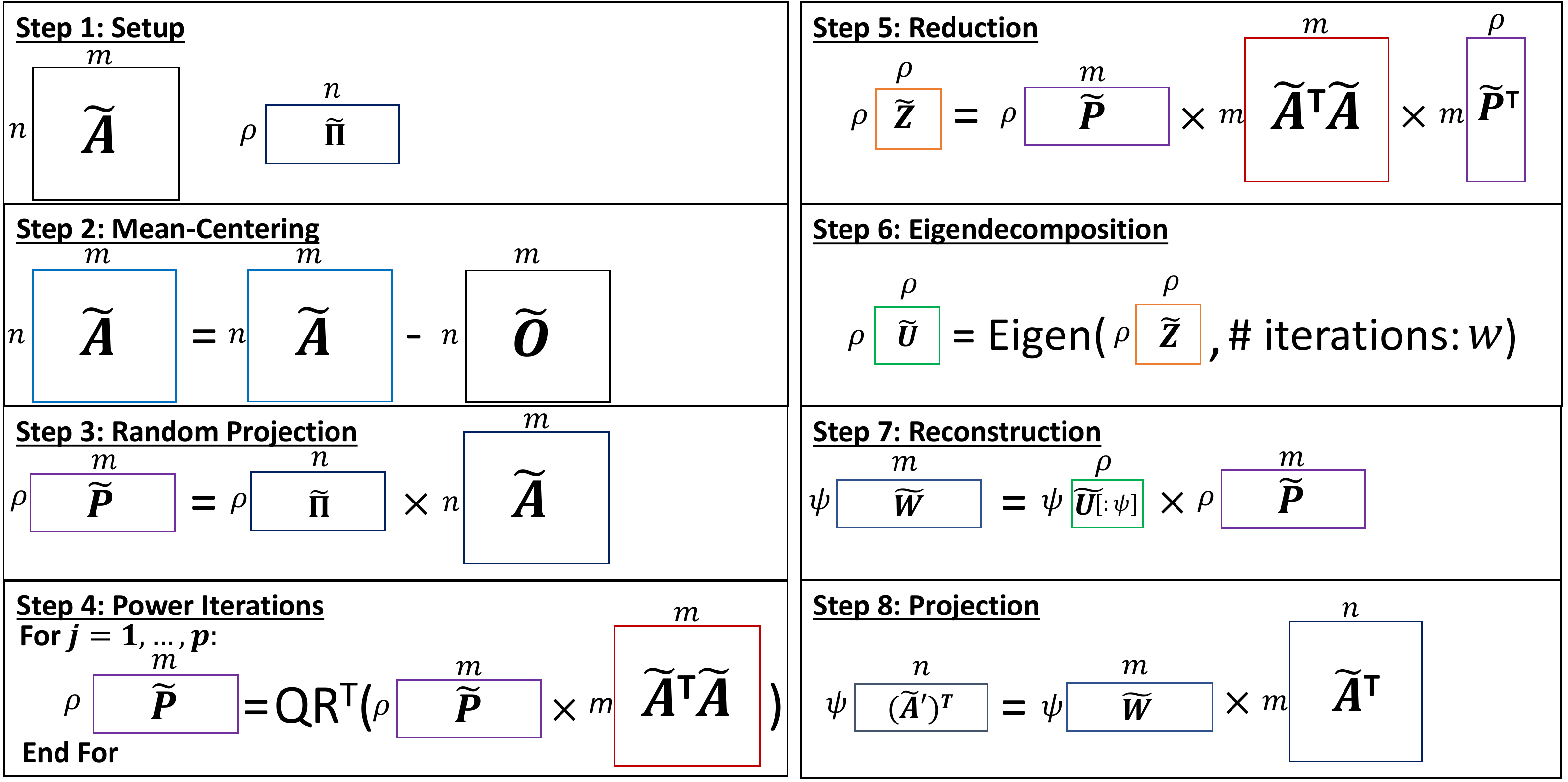}
    \caption{\textbf{Randomized PCA Workflow.} Matrix dimensions are shown with the box sizes and are indicated on the left and top of the corresponding box. 
    }
    \label{fig:randoPCA}
\end{figure}

\noindent \textbf{Randomized PCA (RPCA)}~\cite{halko2011finding} is an efficient randomized algorithm for PCA, which lowers the complexity of the matrix decomposition by first reducing the input dimension via random projection~\cite{halko2011finding}. Fig.~\ref{fig:randoPCA} depicts the workflow of RPCA.
It takes as input the matrix $\bm{\tilde{A}}$ and a random sketch matrix $\bm{\tilde{\Pi}}$ (\textbf{Step 1}). We adopt the count-sketch approach~\cite{charikar2002finding} for generating the latter, where the elements are drawn from $\{-1,0,1\}$. The columns of the input matrix $\bm{\tilde{A}}$ are first mean-centered (\textbf{Step 2}); $\bm{\tilde{O}}$ denotes the matrix in which each column contains the mean of the corresponding column of $\bm{\tilde{A}}$. Next, the input matrix is projected to a lower-dimensional space by multiplying with the sketch matrix (\textbf{Step 3}). 
For improved accuracy~\cite{halko2011finding}, the projected matrix $\bm{\tilde{P}}$ is recursively multiplied with the covariance matrix $\bm{\tilde{A}}^T\bm{\tilde{A}}$ for $p$ iterations (\textbf{Step 4}). At each iteration, the resulting matrix is orthogonalized using the QR factorization for numerical stability. We denote this step by $\text{QR}^T$, as this algorithm is applied to the rows of the matrix, not columns, in our setting. 
In \textbf{Step 5}, a small symmetric matrix $\bm{\tilde{Z}}$ representing the feature covariance in the low-dimensional space is computed by multiplying the result $\bm{\tilde{P}}$ of Step 4 on both sides of the covariance matrix. In \textbf{Step 6}, the eigenvectors $\bm{\tilde{W}}$ of $\bm{\tilde{Z}}$ are computed via eigendecomposition ($\text{Eigen}$); we use the QR iteration algorithm with tridiagonalization and implicit shifting of eigenvalues~\cite{wang2001convergence} (see \S.\ref{subsubsec:matrixtransfo} for details), which are standard techniques for improving the convergence.
RPCA reduces the original problem of factorizing $\bm{\tilde{A}} \in \mathbb{R}^{(n \times m)}$ to decomposing the tiny, constant-size matrix $\bm{\tilde{Z}} \in \mathbb{R}^{(\rho \times \rho)}$, where $\rho = \psi + \alpha$ with $\psi$ the desired number of principal components and $\alpha$ an oversampling parameter. The latter is used to increase the accuracy of the algorithm~\cite{halko2011finding}. In \textbf{Step 7}, it reconstructs the eigenvectors in the original space (i.e., the PCs $\bm{\tilde{W}}$) and finally projects the data points of $\bm{\tilde{A}}$ onto the PCs in \textbf{Step 8} to construct the output.

\noindent \textbf{Multiparty Homomorphic Encryption (MHE).} \label{subsec:mhe}
To securely perform PCA across distributed datasets, we rely on a multiparty (or distributed) fully-homomorphic encryption scheme~\cite{mouchet2019distributedbfv} in which the secret key $sk$ is shared among the parties via a secret-sharing scheme, whereas the corresponding collective public key $pk$ is known to all of them. As a result, each party can independently compute on ciphertexts encrypted under $pk$, but all parties have to collaborate to decrypt a ciphertext.

Mouchet et al.~\cite{mouchet2019distributedbfv} showed how to adapt ring-learning-with-errors-based homomorphic encryption schemes \cite{cheon2017homomorphic,fan2012somewhat,lyubashevsky2010ideal} to the multiparty setting. In \sys, we instantiate the multiparty scheme with the Cheon-Kim-Kim-Song (\textsc{ckks}) cryptosystem~\cite{cheon2017homomorphic}. \textsc{ckks} is a homomorphic encryption scheme that enables approximate arithmetic over $\mathbb{C}^{\mathcal{N}/2}$; the plaintext and ciphertext spaces share the same domain $R_{Q_{L}}=\mathbb{Z}_{Q_{L}}[X]/(X^\mathcal{N}+1)$, with $ Q_{L} = \prod_{0}^{L} q_{i}$ in our case and $\mathcal{N}$ a power of 2. Both plaintexts and ciphertexts are represented by polynomials of degree up to $\mathcal{N}-1$ (with $\mathcal{N}$ coefficients) in this domain\df{, each encoding} a vector of up to $t = \mathcal{N}/2$ floating-point values. Any operation is SIMD, i.e., simultaneously performed on all encoded values. \df{\textsc{ckks}'s security is based on the ring learning with errors (\textsc{rlwe}) problem~\cite{lyubashevsky2010ideal} and some noise is added directly in the least significant bits of the encrypted values. Mouchet et al. \cite{mouchet2019distributedbfv} have shown that the distributed protocols (described below) introduce only additive noise, linear in the number of DPs. To limit the noise growth during homomorphic operations in \sys, we leverage general scale-management techniques for CKKS \cite{costache2022precision, lee2022high, kim2022approximate}.} \df{We refer to Appendix \ref{appendix:mhe} for cryptoscheme details}. 

\emph{Main MHE Operations}. The DPs each have a public key $pk_i$ and the corresponding secret key $sk_i$ (with $\{sk_i\}$ the set of all DPs' secret keys) and can collectively execute the following operations. \df{We denote a collectively encrypted vector by $\bm{c}$ and a plaintext vector by $\bm{\tilde{p}}$. Symbols are summarized in Tab.~\ref{tab:symbols} (Appendix~\ref{app:symbols}).}

\begin{itemize}[leftmargin=*]
    \setlength\itemsep{-0.0em}
    \item $pk, evks \xleftarrow{} \text{DKeyGen}(\{sk_i\})$ generates the collective public key $pk$ and evaluation keys $evks$, which are required for ciphertext transformations such as rotations. \df{The DPs aggregate the local shares of keys (randomly generated based on a public source of randomness) to obtain public collective keys~\cite{mouchet2019distributedbfv}.} 
    \item $\bm{c}$ $\xleftarrow{} \text{DBootstrap}(\bm{c}',\ \{sk_i\})$ collectively refreshes a ciphertext \df{to obtain a fresh encryption. This operation is required after every $\lambda$ multiplications to ensure a correct decryption.}
    \item $\bm{c}_{pk'} \xleftarrow{} \text{DKeySwitch}(\bm{c},\ pk',\ \{sk_i\})$ changes the encryption of a ciphertext $\bm{c}$ from the public key $pk$ to another public key $pk'$, without decrypting the ciphertext. The collective decryption is a special case of this operation (i.e., $\text{DKeySwitch}(\bm{c},\emptyset,\{sk_i\})$). \df{To prevent information leakage upon decryption \cite{li2021security}, a fresh noise with a variance larger than that of the ciphertext is added before decryption \cite{mouchet2019distributedbfv, li2021security, cheon2020remark,de2020fast}}.
\end{itemize}

\noindent Each DP can independently encrypt, and perform the following operations listed in order of increasing computational complexity (Tab.~\ref{tab:localruntimes}):

\begin{itemize}[leftmargin=*]
    \setlength\itemsep{-0.0em}
    \item $\bm{c}_{pk}\in R^2_{Q_{L}} \xleftarrow{} \text{Enc}(pk, \bm{\tilde{p}})$ with a plaintext vector $\bm{\tilde{p}}$, such that $\text{DKeySwitch}(\bm{c}_{pk}, \emptyset, \{sk_i\}) \approx \bm{\tilde{p}}$.
    \item $\bm{c}'' =\bm{c}$ + $\bm{c}'$, addition of encrypted vectors.
    \item $\bm{c}' = \bm{c} \cdot \bm{\tilde{p}}$, element-wise multiplication of an encrypted vector and a cleartext vector. The result needs to be rescaled to maintain ciphertext scale.
    \item $\bm{c}'' = \bm{c} \cdot \bm{c'}$, element-wise multiplication of two encrypted vectors. The result needs to be relinearized and rescaled to maintain ciphertext size and scale.
    \item $\bm{c}' = \text{Rot}_y(\bm{c},evks)$, cyclic rotation of length $y$ to the left (to the right if $y$ is negative) on the encrypted vector $\bm{c}$.
    \item $\bm{c}'' = \bm{c} \bullet \bm{c}'$, dot product of two encrypted vectors. The result is encoded in the first position of a one-hot encoded vector $\bm{c}''$. 
    \item $\bm{c}' = \text{Dup}_y(\bm{c})$, duplication of the first element of $\bm{c}$ to the first $y$ positions of $\bm{c}'$ with $\lceil \log_2(y) \rceil$ rotations and additions. 
\end{itemize}
\section{\df{SF-PCA} System and Security Models }\label{sec:overview}

Our system model is illustrated in Fig.~\ref{fig:model}. We consider a cleartext dataset, represented as a matrix $\bm{\tilde{A}} \in \mathbb{R}^{(n \times m)}$, that is horizontally split among a set of interconnected data providers $\text{DP}_1,\dots,\text{DP}_{s}$ such that each DP$_i$ has $\bm{\tilde{A}}_i \in \mathbb{R}^{(n_i \times m)}$ with $\sum_i n_i = n$. 
The number of data samples held by each DP (i.e., $n_i$) is considered public.
We discuss the vertically partitioned case in \S.\ref{sec:extensions}. 
\begin{figure}[ht]
    \centering
    \footnotesize
    \includegraphics[width=1.0\columnwidth]{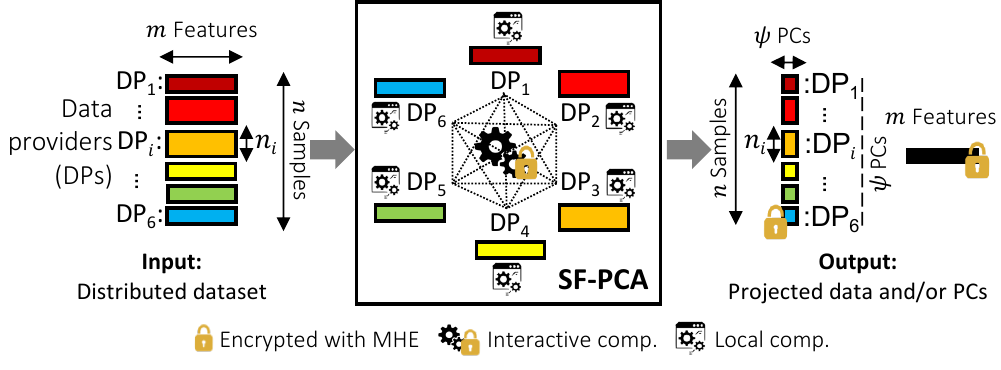}
    \vspace{-1.0em}
    \caption{{\textbf{\sys System Model and Functionality.} Each DP$_i$ holds a $n_i \times m$ submatrix as input and collaboratively executes \sys to obtain encrypted PCs and/or the projection of its local data.}}
    \label{fig:model}
\end{figure}

\sys enables the DPs to collaboratively execute a randomized PCA on their joint data. In the end, each DP obtains $\psi$ collectively encrypted PCs, on which each DP can locally project its data. If required by the application, each DP's projected data (encrypted under the collective key) can be collectively switched ($\text{DKeySwitch}$, \S.\ref{subsec:mhe}) to each DP's public key $pk_i$ to be locally decrypted. Similarly, the PCs can be collectively decrypted and shared among the DPs.

We adopt the semi-honest model, where the DPs follow the protocol as specified, but might try to infer information about another DP's data, potentially colluding with other DPs.
We require that the DPs' data and all intermediate results remain confidential. In other words, \sys provides \textit{input confidentiality}, i.e., no DP is able to learn any information about any other DP's local data other than what it can infer from the final output of PCA (e.g., its projected local data).
We require that this property holds as long as one DP remains honest and does not collude with others.

\section{\df{SF-PCA} Protocol Design}\label{sec:design}

We introduce an end-to-end confidential and federated approach to execute a RPCA (\S.\ref{subsec_randoPCA}) jointly over $s$ DPs holding their local data. At each step of the PCA execution, the DPs collectively compute encrypted global intermediate results through interactive protocols that combine the results of local computation on each DP's cleartext data. The intermediate results remain encrypted under the DPs' collective key and are never revealed. While our system's ability to leverage local cleartext computation opens the door to efficient multiparty algorithms, a careful algorithmic design is still necessary for developing a practical PCA protocol. 

Leveraging existing approaches for secure computation (e.g., HE or SMC), the DPs could outsource their encrypted (or secret-shared) data to one or multiple computing parties to jointly perform the PCA. However, the communication overhead of sharing the entire dataset as well as the computational burden of performing complex computations (e.g., multiplication and factorization of matrices) on the pooled dataset render these solutions impractical for large-scale datasets. Note that the repeated matrix multiplications are challenging to perform efficiently under HE due to the costly bootstrapping procedure. 
\sys addresses these challenges by introducing efficient MHE-based protocols based on a federated approach to joint computation.
We compare \sys's performance with existing approaches in \S.\ref{sec:eval}. 

\subsection{
\df{Key Strategies for Accuracy and Efficiency} 
}
\label{subsec:optimizations}
\df{In RPCA (Fig.~\ref{fig:randoPCA}), many matrix multiplications involving the input matrix $\bm{\tilde{A}}^{(n \times m)}$ (or its covariance matrix) largely determine the protocol's complexity.} These multiplications are interspersed with sophisticated linear algebra transformations, such as the QR factorization invoked at the end of each power iteration (Step 4 in Fig.~\ref{fig:randoPCA}) and eigendecomposition (Eigen in Step 6), which view the matrix as a set of row (or column) vectors and apply vector-level operations. Below, we explain our strategies \df{to carry out these computations efficiently while maintaining the accuracy of results.}

\noindent \df{\textbf{Obtaining Accurate Results by Emulating Centralized PCA.} Existing federated approaches to PCA that combine the results independently obtained by the DPs (e.g., meta-analysis), are prone to errors introduced by differences in data distribution among the DPs. In \sys, we  avoid this pitfall by securely combining the intermediate results at each step of the protocol (via collective aggregation $\Xi$ in Alg.\ref{alg:workflow})
to emulate a centralized analysis, thus obtaining the same PCs regardless of how the data are split (\S.\ref{subsec:evalAppl}).
}

\noindent \textbf{Efficient Edge-Computing on Local Cleartext Data.}
\df{Working with an encrypted form of the entire input matrix ($\bm{\tilde{A}}$ in Fig.~\ref{fig:randoPCA}) would require the DPs to transfer large amounts of data (e.g., for centralized HE or secret sharing) or to perform costly ciphertext operations on large matrices, both of which become impractical for large-scale datasets. 
In \sys (Alg.\ref{alg:workflow}),
the DPs jointly perform the PCA without encrypting or exchanging the input data.
They collaborate instead by computing on their local cleartext data (i.e., the sub-matrix $\bm{\tilde{A}}_i$) and exchanging only low-dimensional and aggregate-level encrypted information.} 
This enables the DPs to minimize communication and maximize the use of low-cost MHE operations involving the cleartext data (e.g., with our default parameters, cleartext-ciphertext multiplication is eight times faster than a ciphertext-ciphertext multiplication; Tab.~\ref{tab:localruntimes}). We also modify the RPCA computation to use only the cleartext input throughout the workflow. \df{For example, instead of directly constructing a mean-centered input matrix (Step 2 in Fig.~\ref{fig:randoPCA}), which needs to be encrypted due to the means being private, \sys keeps each local matrix $\bm{\tilde{A}}_i$ in cleartext and associates with it an encrypted mean vector $\bm{o}$ to correct for mean shifts in subsequent steps (see Step 2 in Alg.\ref{alg:workflow}).} \df{This enables a key optimization for the matrix multiplications in Steps 3-5, 7 and 8 (Alg.\ref{alg:workflow})}, where the cleartext matrix $\bm{\tilde{A}}_i$ is pre-transformed to minimize costly ciphertext operations such as rotations in later steps. \df{In \S.\ref{subsub:dupvecmatmult}, we show how to efficiently multiply an encrypted matrix with another containing only duplicated rows (or columns), which is used for lazy mean correction in Steps 4, 5, 7 and 8.}

\noindent \textbf{Adaptive Selection of Computational Routines based on Data Dimensions.}
\df{In practice, PCA is applied to datasets whose dimensions vary greatly depending on the application, e.g., from tens of features in small predictive modeling tasks to tens of thousands of features in genomic studies (\S.\ref{sec:eval}). To achieve practical performance in a wide range of settings, we propose an adaptive approach for optimizing the computational routines based on the input dimensions. In Alg.\ref{alg:workflow}, we introduce two different workflows for performing RPCA: \textit{Precomp} and \textit{Seq}. In \textit{Precomp}, the encrypted covariance matrix $\bm{G}$ is precomputed in the beginning of Step 4 and reused, such that most of the following operations scale primarily with the number of features $m$. In \textit{Seq}, $\bm{\tilde{A}}$ is kept in cleartext and used for matrix multiplications, which is more efficient than using $\bm{G}$, but now the computation scales with both $m$ and the number of samples $n$.
In addition, in \S.\ref{subsub:matmult}, we describe several matrix multiplication methods, each of which scales differently with the input dimensions; \sys selects the best approach for each step in its workflow.
Similarly, in \S.\ref{subsubsec:matrixtransfo}, we introduce two approaches for performing the QR factorization on an encrypted matrix (QR in Step 4), with different complexities depending on the input dimensions.}

\noindent \textbf{Optimized Data Encoding for  Linear Algebra  on Encrypted Matrices.}
The \df{secure execution of} RPCA requires that the DPs iteratively perform various matrix operations on encrypted data, including multiplication and factorization. \df{For example, the QR factorization, which is repeatedly executed in-between matrix multiplications (in Steps 4, 6, and 7 of RPCA; Fig.~\ref{fig:randoPCA}), is performed over the rows of an encrypted matrix in \sys. Selecting a row in a matrix of $m$ columns, where the columns are individually packed in ciphertexts, would require $m$ homomorphic multiplications, additions, and rotations; in contrast, row selection incurs no cost when the matrix is row-wise encoded. In fact, the overwhelming cost of transforming encrypted matrices from one encoding to another would make our system impractical.} We \df{therefore} adopt a consistent vectorized encoding scheme throughout the algorithm to represent encrypted matrices and tailor the operations to efficiently work with this format without costly conversions. This also allows \sys to fully utilize the packing and SIMD properties of MHE thus minimizing its overall computation and communication costs.

\noindent \textbf{Selective Bootstrapping to Minimize Communication.}
After a certain number of multiplications, a ciphertext needs to be bootstrapped ($\text{DBootstrap}$, \S.\ref{subsec:mhe}) to restore its capacity for computation. In \sys, this is a collective operation, which is computationally lightweight in contrast to its centralized equivalent, but requires the ciphertext to be exchanged among all DPs. To further minimize this communication overhead, we restrict the invocation of $\text{DBootstrap}$ to places where an intermediate result is already globally synced and of a small dimension (e.g., during QR factorization \df{in Steps 4 and 7 in Alg.~\ref{alg:workflow}}; see \S.\ref{subsubsec:matrixtransfo}), while flexibly allowing a ciphertext to be bootstrapped even if some multiplication capacity remains.

\subsection{Workflow Details} \label{subsec:workflow}
We describe the workflow of \sys from the point of view of DP$_i$ in Alg.~\ref{alg:workflow}. Recall that the DPs aim to compute $\psi$ encrypted PCs (rows of matrix $\bm{W}$) on their joint data.
RPCA identifies $\rho = \psi + \alpha$ components with a small oversampling parameter $\alpha$ for improved accuracy. In addition to Alg.~\ref{alg:workflow}, we show in Fig.~\ref{fig:randoPCA_Verti} in the Appendix how the matrix dimensions evolve in \sys's workflow.
The DPs interact by aggregating (represented by $\Xi$)
encrypted matrices and broadcasting the encrypted result to all DPs.

 \begin{figure}[!ht]
 \setcounter{figure}{0}
     \centering
     \small
     \includegraphics[width=\columnwidth]{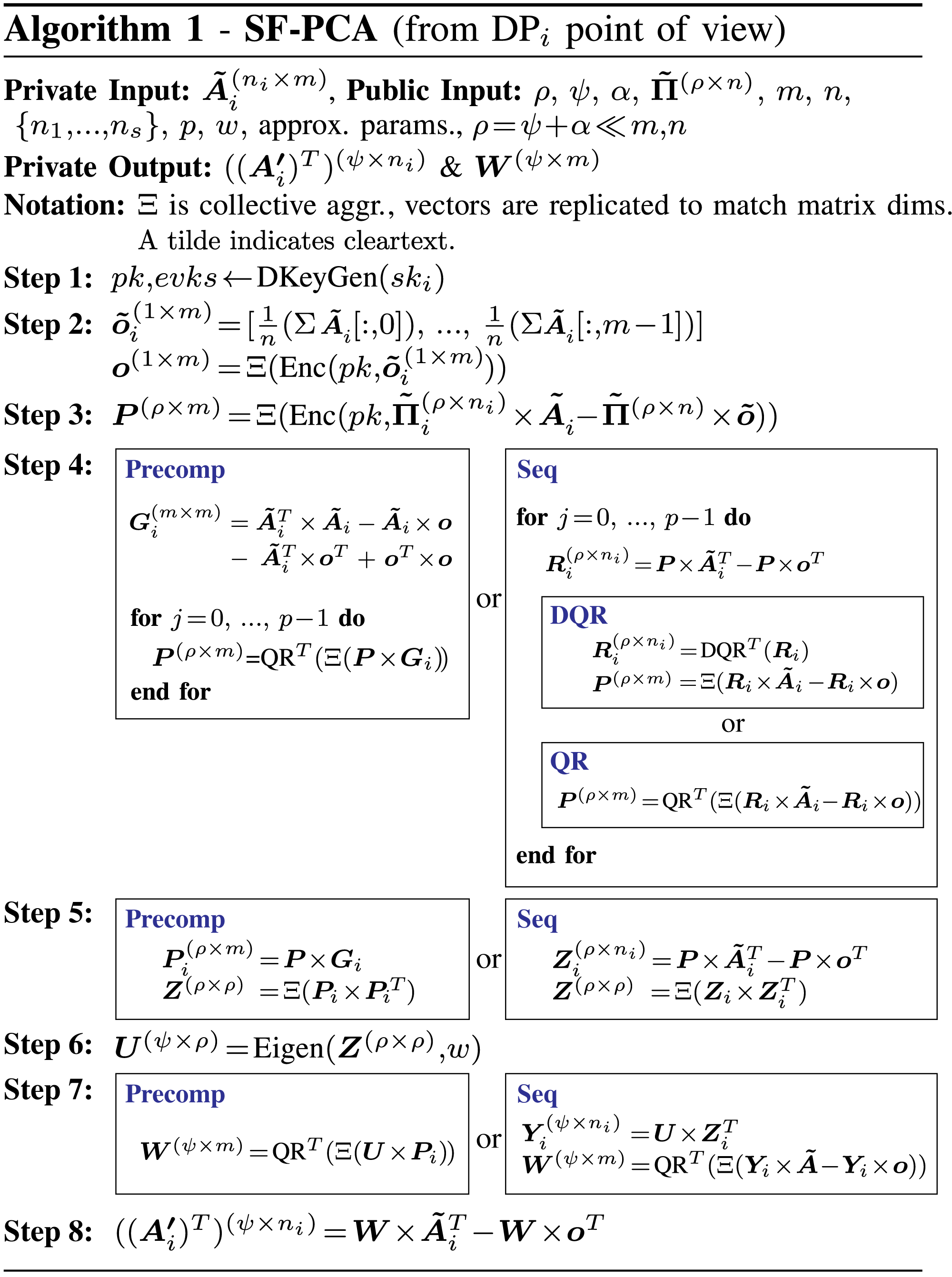}
     \vspace{-0.75em}
    \captionsetup{labelformat=empty}
    \caption{}
     \label{alg:workflow}
     \setcounter{figure}{2}
     \vspace{-2.0em}
 \end{figure}

\noindent\textbf{Step 1: Setup.} Each DP$_i$ holds $\bm{\tilde{A}}^{(n_i \times m)}$, a submatrix of the global input matrix $\bm{\tilde{A}}^{(n \times m)}$. The DPs \df{generate the required public keys (\df{DKeyGen}, \S.\ref{subsec:mhe}) and }agree on the PCA parameters, including: the number of power iterations $p$, the number of QR iterations $w$ for eigendecomposition, the desired number of PCs $\psi$, the oversampling parameter $\alpha$ (resulting in the number of components $\rho=\psi+\alpha$ for RPCA), and a public random sketch matrix $\bm{\tilde{\Pi}}^{(\rho \times n)}$ (e.g., generated from a shared seed).
In addition, the DPs together decide the specifics of certain computational steps in \sys, such as the approximation intervals for non-linear operations (\S.\ref{subsubsec:approx}) and the method of choice for costly linear algebraic operations (matrix multiplication and transformations; see \S.\ref{subsec:mheops}), taking the input dimensions into account to maximize performance. 
All the parameters introduced in this step are considered public.
Note that the procedure to agree upon the parameters is orthogonal to \sys; e.g., the DP initiating the collaboration could propose the parameters.

\noindent\textbf{Step 2: Mean Calculation.} The DPs compute the encrypted vector $\bm{o}^{(1 \times m)}$ of column averages of the input matrix $\bm{\tilde{A}}^{(n \times m)}$ by securely aggregating their local column sums divided by $n$, encrypted under the collective public key. 

\noindent\textbf{Step 3: Random Projection.} DP$_i$ projects $\bm{\tilde{A}}_i$ to a subspace of $\rho$ dimensions using the public sketch matrix $\bm{\tilde{\Pi}}^{(\rho \times n)}$. DP$_i$ locally computes the product of  $\bm{\tilde{A}}_i$ and the corresponding submatrix $\bm{\tilde{\Pi}}_i^{(\rho \times n_i)}$ of $\bm{\tilde{\Pi}}^{(\rho \times n)}$ to obtain its local sketch in cleartext. The result is then encrypted and aggregated among all DPs to obtain the encrypted sketch $\bm{P}$ of the global matrix $\bm{\tilde{A}}$.  

\noindent\textbf{Step 4: Power Iterations.} 
The sketch of the input matrix obtained in the previous step is repeatedly multiplied with the input matrix to increase the spectral gap between the top eigenvectors of interest and the rest~\cite{halko2011finding}.
We execute this step differently depending on the input dimensions for optimized performance; the two approaches considered by \sys are described below.
Notably, in both approaches, we leverage the fact that the cost of cleartext operations is almost negligible compared to that of HE to optimize the computation.

\textit{Approach 1: Precompute \& Reuse} (\textsl{Precomp}): Each DP$_i$ precomputes the covariance matrix $\bm{G}_i^{(m \times m)}$ once and reuses it in every iteration for multiplying with $\bm{P}$.
Note that $\bm{G}_i$ needs to be encrypted due to the mean-centering operation using the encrypted global column means $\bm{o}^{(1 \times m)}$. \sys's optimized matrix multiplication routines between an encrypted and a cleartext matrix (\S.\ref{subsub:matmult}) minimize the computation involving the encrypted matrix by precomputing certain transformations of the cleartext matrix at a negligible cost. To efficiently apply these methods to the encrypted $\bm{G}_i$, we obtain the transformations of $\bm{G}_i$ by transforming the cleartext $\bm{\tilde{A}}_i$ and $\bm{\tilde{A}}_i^T$ before multiplying them.

\textit{Approach 2: Sequentially Multiply} (\textsl{Seq}): The DPs sequentially multiply $\bm{P}$ by the cleartext matrix $\bm{\tilde{A}}_i$ (and its transpose) on the fly. The covariance matrix is never explicitly constructed.
To keep the input matrix $\bm{\tilde{A}}_i$ in cleartext, \sys performs the mean-centering of $\bm{\tilde{A}}_i$ in a \textit{lazy} manner (\textit{lazy mean-centering}): instead of subtracting the encrypted vector $\bm{o}^{(1 \times m)}$ from each row of $\bm{\tilde{A}}_i$, which would transform the whole input matrix into an encrypted matrix, multiplication is performed using the original cleartext $\bm{\tilde{A}}_i$ and the resulting matrix is corrected to account for the mean shift. More precisely, we multiply encrypted $\bm{P}$ with mean-centered $\bm{\tilde{A}}_i$ in three efficient steps: (1) multiply $\bm{P}$ with the cleartext matrix $\bm{\tilde{A}}_i$, (2) compute the inner product between each row of $\bm{P}$ and $\bm{o}$ (see \S.\ref{subsub:dupvecmatmult}), and (3) subtract each inner product value from all elements in the corresponding row of the matrix from (1).
\df{We observe that \textit{Precomp} requires fewer multiplications per power iteration and its computation cost is mostly independent of $n$. \textit{Seq} requires more operations but maximizes cleartext operations by reusing the cleartext matrix $\bm{\tilde{A}}$. We compare the performance of both approaches in \S.\ref{sec:eval}.}

In each iteration, a QR factorization (QR$^T$; Alg.~\ref{alg:qrt}) is applied to either the aggregated matrix $\bm{P}^{(\rho \times m)}$, in both approaches, or the intermediate $(\bm{P} \times \bm{\tilde{A}}^T_i)^{(\rho \times n_i)}$ in $\textsl{Seq}$.
In the latter, the factorization is optionally performed using a new interactive protocol DQR$^T$, when the computational speedup of each DP computing on a matrix with $n_i$ columns vs. one DP computing on an aggregated matrix with $m$ columns exceeds the additional communication cost, i.e., when $n_i \ll m$ (see \S.\ref{subsubsec:matrixtransfo}).

\noindent\textbf{Step 5: Reduction.} In the \textsl{Precomp} approach, the matrix $\bm{P}$ resulting from Step 4 is transformed to a small symmetric matrix $\bm{Z}$ by multiplying the covariance matrix $\bm{G}_i$ from both sides. In the \textsl{Seq} approach, this is performed by using the cleartext matrix $\bm{\tilde{A}}_i^T$ and then by multiplying the result by its transpose. As in Step 4, \sys employs \textit{lazy mean-centering} for this step.

\noindent\textbf{Step 6: Eigendecomposition.} The eigendecomposition (introduced in \S.\ref{subsec_randoPCA}) is executed on the encrypted matrix $\bm{Z}$. We detail our MHE-based algorithm for this step in Alg.~\ref{alg:eigendecompo}. It requires the iterative execution of QR$^T$ and matrix multiplications. 

\noindent\textbf{Step 7: Reconstruction.} The PCs (rows of $\bm{W}$) are computed by multiplying the eigenvectors from Step 6 with the approximated subspace from Step 3, followed by a final round of power iteration and orthogonalization (QR$^T$) for numerical stability.

\noindent\textbf{Step 8: Projection.} Each DP$_i$ projects its local cleartext data $\bm{\tilde{A}}^T_i$ onto the collectively encrypted PCs in $\bm{W}$ to obtain their projected data $\bm{A'}_i$, which is also encrypted under the collective public key. If required by the application, by using $\text{DKeySwitch}$, the PCs and/or the DPs' projected data can be collectively decrypted or re-encrypted under the public keys of specific entities to grant controlled access to the decrypted results.

\section{Optimized Routines for Linear Algebra and Non-Polynomial Functions on Encrypted Data} \label{subsec:mheops}
We describe how \sys efficiently executes matrix multiplications, sophisticated linear-algebra transformations and non-polynomial function evaluations on encrypted data. 
\df{Although the methods in this section can also be employed in the centralized setting, we note that the adaptive use of matrix multiplication routines and the higher-level protocols for matrix transformations (e.g., QR factorization) are optimized while accounting for the unique properties of MHE, e.g., the availability of local cleartext data and a lightweight interactive bootstrapping routine, both of which alter the tradeoff between different computational strategies and present new ways to optimize the algorithm. Our secure federated routines may be of independent interest for other applications.} 

\subsection{Matrix Multiplications}\label{subsub:matmult}

\df{Encrypted matrix multiplications are frequently invoked in \sys's workflow and hence are a key determinant of its performance. As outlined in Alg.~\ref{alg:workflow}, we introduce two high-level algorithmic workflows---\textsl{Precomp} and \textsl{Seq}---for executing RPCA. Both approaches involve different types of multiplications over matrices of varying dimensions, motivating our adaptive strategy for choosing the most efficient routine for each computational step in \sys among a range of multiplication methods.}

\subsubsection{\df{Adaptive Strategy}}\label{subsub:adaptstrat}
\df{We identify two main types of matrix multiplications in Alg.~\ref{alg:workflow}: (i) \textit{unbalanced multiplications} between a large encrypted matrix and a large cleartext (or pre-transformed encrypted) matrix in Steps 4, 7 and 8, with the key property that operations are cheap on one matrix (cleartext) and expensive on the other (ciphertext); 
and (ii) \textit{duplicated-vector multiplications}, referring to multiplications between a large encrypted matrix and another encrypted matrix whose rows (or columns) are identical (e.g., corresponding to the encrypted mean vector $\bm{o}$).

In \S.\ref{subsub:unbalmatmult}, we detail three different approaches (M\ref{alg:m2}, M\ref{alg:m3}, and M\ref{alg:m1}) for unbalanced multiplications, each with a complexity that scales differently with the input dimensions.} We denote by $\zeta^*_{\text{M}}(a,b,c)$, the function that takes the three input dimensions for multiplying $\bm{M}^{(a\times b)}$ and $\bm{\tilde{N}}^{(b\times c)}$ matrices and outputs the cost associated with the most efficient multiplication routine. The cost we compare is a weighted sum of the multiplication and rotation invocation counts, where the weights are determined by the estimated runtime per operation in the given computational environment. \df{The cost of a cleartext-ciphertext multiplication is set to be around 8 times lower than that of a ciphertext-ciphertext multiplication according to our estimates (Tab.~\ref{tab:localruntimes}).}

We identify the matrix multiplication costs of \textsl{Precomp} and \textsl{Seq} for a single iteration as $\zeta^*_{\text{M}}(\rho,m,m)$ and $\zeta^*_{\text{M}}(\rho, m, n_{\max} ) + \zeta^*_{\text{M}}(\rho, n_{\max} , m)$, respectively, where $\rho$ is the number of reduced dimensions in RPCA, $m$ is the number of input features, and $n_{\max} = \max_i (n_i)$ represents the largest number of samples locally held by the DPs.
We consider the worst-case complexity as the overall runtime being as fast as the slowest DP.
In addition, for both approaches, we incorporate the cost of lazy mean-centering when comparing the overall cost; \textsl{Precomp} requires $3b$ Mults and $b$ Rots, whereas \textsl{Seq} requires two \textit{duplicated-vector multiplications}, which we detail in \S.\ref{subsub:dupvecmatmult}. We further compare these approaches in \S.\ref{sec:eval}.
\subsubsection{\df{Unbalanced Multiplications}}\label{subsub:unbalmatmult}
We describe the HE implementations of three matrix multiplication strategies:  Dot-Product Method (M\ref{alg:m2}), Element-Duplication Method (M\ref{alg:m3}), and Diagonal Method (M\ref{alg:m1}; adapted from Jiang et al. \cite{jiang2018secure}). We jointly consider these three methods because their costs scale differently with the input dimensions, enabling \sys to optimize its performance in a wide range of scenarios. For each method, we show its cost in terms of the invocations of ciphertext rotations (Rots) and multiplications (Mults) for multiplying a pair of $a\times b$ and $b\times c$ matrices. The cost of cleartext operations is negligible. To simplify the computational complexity analysis, we assume that $b$ and $c$ are powers of two without loss of generality.
We denote by $t$ the ciphertext capacity, i.e., the number of values that can be packed in a ciphertext. \df{Due to \sys's vectorized encoding, the inner dimension $b$ reduces to a small constant $\ceil{\frac{b}{t}}$ in terms of the number of ciphertext operations.}

\noindent\textbf{Dot-Product Method \df{(M1)}.} Each element of $\bm{R}$ is obtained from the dot product ($\bullet$) between a row of $\bm{M}$ and a column of $\bm{\tilde{N}}$ (Line~4 in M\ref{alg:m2}). The result of the dot-product is moved to position $j$ (0-based) by masking and rotating the vector by $j$ positions to the right (i.e., $\text{Rot}_{(-j)}$; see \S.\ref{subsec:mhe}). In \sys, this method is used (in Step~5 in Alg.~\ref{alg:workflow}) to multiply an encrypted matrix by its transpose without any additional transformation, since the encrypted rows of $\bm{M}$ can be directly used as the columns of $\bm{N}$. \df{The multiplication and rotation costs mainly depend on the outer dimensions.
}
\begin{method}{\textbf{: Dot-Product Method}}
\footnotesize
	\setcounter{method}{\value{method}-1}
	\refstepcounter{method}\label{alg:m2}
	\renewcommand{\themethod}{}
	\begin{algorithmic}[1]
	\vspace{-0.75em}
	\item[\textbf{Input:} Encrypted $\bm{M}^{(a \times b)}$ and cleartext \df{(indicated by a tilde)} $\bm{\tilde{N}}^{(b \times c)}$.]
	\item[\textbf{Output:} Encrypted $\bm{R}^{(a \times c)} = \bm{M} \times \bm{\tilde{N}}$]
	\item[\textbf{Cost:} $(\ceil{\frac{b}{t}}+1) \cdot a c$ Mults and $a c \cdot \ceil{\frac{b}{t}} \cdot \log_2(t)$ Rots]  \
	\STATE $\bm{R} \leftarrow \textbf{0}^{(a\times c)}$
	\FOR{$i = 0,\ \dots,\ a-1$}
	    \FOR{$j = 0,\ \dots,\ c-1$}
	        \STATE $\bm{R}[i,:] \leftarrow \bm{R}[i,:] + \text{Rot}_{(-j)}(\bm{M}[i,:] \bullet \bm{\tilde{N}}[:,j])$
	    \ENDFOR
	\ENDFOR
	\vspace{-1.0em}
	\end{algorithmic}
\end{method}

\noindent\textbf{Element-Duplication Method \df{(M2)}.} \df{This method avoids the computation of pairwise dot products (used in M\ref{alg:m2})} by duplicating each element of $\bm{M}$ to construct a vector of length $c$ and by \df{multiplying} this vector (element-wise) with \df{each} row of $\bm{\tilde{N}}$ (Line~4 in M\ref{alg:m3}). The results are aggregated to obtain $\bm{R}$. \df{This method's cost depends mostly on the left matrix dimensions. 
}
\begin{method}{\textbf{: Element-Duplication Method}}
	\setcounter{method}{\value{method}-1}
	\refstepcounter{method}\label{alg:m3}
	\renewcommand{\themethod}{}
	\footnotesize
	\begin{algorithmic}[1]
	\vspace{-0.75em}
	\item[\textbf{Input:} Encrypted $\bm{M}^{(a \times b)}$ and cleartext \df{(indicated by a tilde)} $\bm{\tilde{N}}^{(b \times c)}$]
	\item[\textbf{Output:} Encrypted $\bm{R}^{(a \times c)} = \bm{M} \times \bm{\tilde{N}}$]
	\item[\textbf{Cost:} $\ceil{\frac{b}{t}} \cdot a  b$ Mults and $a b \cdot \log_2(\min\{c,t\}) $ Rots]
	\STATE $\bm{R} \leftarrow \textbf{0}^{(a\times c)}$
	\FOR{$i = 0,\ \dots,\ a-1$}
	    \FOR{$j = 0,\ \dots,\ b-1$}
	        \STATE $\bm{R}[i,:] \leftarrow \bm{R}[i,:] + (\text{Dup}_c(\bm{M}[i,j]) \cdot \bm{\tilde{N}}[j,:])$
	    \ENDFOR
	\ENDFOR
	\vspace{-1.0em}
	\end{algorithmic}
\end{method}

\noindent\textbf{Diagonal Method \df{(M3)}.} This approach is based on the technique of Jiang et al.~\cite{jiang2018secure}, which we adapt to large-scale matrices that cannot be packed in a single ciphertext. This method transforms the cleartext matrix (by rotating its columns) such that one of its rows corresponds to the diagonal of the original matrix (Line 2 in M\ref{alg:m1}). The rows of the encrypted $\bm{M}$ are then rotated (Line 8) before being multiplied with the transformed rows of $\bm{\tilde{N}}$ at each iteration along the common dimension $b$ (Line 9). We use the baby-step giant-step approach \cite{shanks1971class} to reduce the number of rotations on the rows of $\bm{M}$ from $b$ to $2 \sqrt{b}$ by storing the intermediate results in three-dimensional tensors (i.e., $\bm{M}'$ and $\bm{R}'$) (Lines 8 and 9), \df{introducing a tradeoff between computation and memory usage (see \S.\ref{subsec:scalability})}. The intermediate results are then aligned and aggregated in the final matrix $\bm{R}$ (Line 14).
The rows of $\bm{M}$ are duplicated or truncated to have $c$ elements ($\text{Len}_c$($\cdot$)) before the multiplication. 
\df{This method's cost also depends mostly on the dimension of the left matrix but, contrary to M\ref{alg:m3}, its number of rotations scales with the square root of the inner dimension times the number of packed ciphertexts along the same dimension.}
\begin{method}{: \textbf{Diagonal Method}}
\footnotesize
    \setcounter{method}{\value{method}-1}
	\refstepcounter{method}\label{alg:m1}
	\renewcommand{\themethod}{}
	\begin{algorithmic}[1]
	\vspace{-0.75em}
	\item[\textbf{Input:} Encrypted $\bm{M}^{(a \times b)}$ and cleartext  \df{(indicated by a tilde)} $\bm{\tilde{N}}^{(b \times c)}$]
	\item[\textbf{Output:} Encrypted $\bm{R}^{(a \times c)} = \bm{M} \times \bm{\tilde{N}}$]
	\item[\textbf{Cost:} $\ceil{\frac{b}{t}} \cdot a b$ Mults and $\ceil{\frac{b}{t}} \cdot (\ceil{\frac{c}{b}} + 2 a \cdot \lceil \sqrt{b} \rceil)$ Rots]
	\FOR{$i = 0,\ \dots,\ c-1$}
	\STATE $\bm{\tilde{N}}[:,i] \leftarrow \text{Rot}_i(\bm{\tilde{N}}[:,i])$ 
	\ENDFOR
	\STATE $\bm{M'} \leftarrow \textbf{0}^{(a\times \ceil{\sqrt{b}}\times c)},\ \bm{R'} \leftarrow \textbf{0}^{(a\times \ceil{\sqrt{b}}\times c)}$
	\FOR{$i = 0,\ \dots,\ b-1$}
	    \STATE $y \leftarrow i\ \text{mod}\ \lceil \sqrt{b} \rceil$,\  $g \leftarrow \lfloor i/\lceil \sqrt{b} \rceil \rfloor$
	    \FOR{$j =\ 0,\ \dots,\ a-1$}
	        \STATE \textbf{if} $\bm{M}'[j,y,:] = \emptyset$ \textbf{then} $\bm{M}'[j,y,:] \leftarrow \text{Len}_c(\text{Rot}_{y}(\bm{M}[j,:]))$
	        \STATE $\bm{\bm{R}}'[j,g,:] \leftarrow \bm{\bm{R}}'[j,g,:] + \bm{M}'[j,y,:] \cdot  \bm{\tilde{N}}[i\ \text{mod}\ m,:]$
	    \ENDFOR
	\ENDFOR
	\FOR{$i = 0,\ \dots,\ a-1$}
	    \FOR{$l = 0,\ \dots,\ \lceil \sqrt{b} \rceil -1$} 
	        \STATE $\bm{R}[\df{i},:] \leftarrow \bm{R}[\df{i},:] + \text{Rot}_{l \cdot \lceil \sqrt{b} \rceil} (\bm{R}'[i,l,:])$
	    \ENDFOR
	\ENDFOR
	\vspace{-1.0em}
	\end{algorithmic}
\end{method}


\subsubsection{\df{Duplicated-Vector Multiplications}}\label{subsub:dupvecmatmult} 

This method addresses a special setting where we multiply an encrypted matrix $\bm{M}$ with another encrypted matrix $\bm{\Gamma}$ whose rows (\textit{Case 1}) or columns (\textit{Case 2}) are identical vectors $\bm{\mu}$.
This setting frequently arises in \sys for the \textit{lazy mean-centering} operations (\df{i.e., all operations involving $\bm{o}$ in} \S.\ref{subsec:workflow}).
Our method accounts for this redundancy in the matrix to minimize the number of rotations on both encrypted matrices.
\begin{method}{\textbf{: Vector-Duplication Method}} 
	\setcounter{method}{\value{method}-1}
	\refstepcounter{method}\label{alg:mv}
	\renewcommand{\themethod}{}
	\footnotesize
	\begin{algorithmic}[1]
	\vspace{-0.75em}
	\item[\textbf{Input:} Encrypted $\bm{M}^{(a \times b)}$ and encrypted $\bm{\Gamma}^{(b \times b)}$,]
	\item[\ \ \ \ \ \ \ \ \ \ \ \,where $\bm{\Gamma}=\textbf{1}\times \bm{\mu}^T$ (\textit{Case 1}) or $\bm{\Gamma}=\bm{\mu}\times \textbf{1}^T$ (\textit{Case 2})]
	\item[\textbf{Output:} Encrypted $\bm{R}^{(a \times b)} = \bm{M} \times \bm{\Gamma}$]
	\item[\textbf{Cost:} $\ceil{\frac{b}{t}} \cdot a$ Mults and $\ceil{\frac{b}{t}} \cdot 2a \cdot \log_2(\min\{b,t\})$ Rots]
	\STATE $\bm{R} \leftarrow \textbf{0}^{(a\times b)}$
	\FOR{$i = 0,\ \dots,\ a$}
	    \STATE \textbf{if} \textit{Case 1} \textbf{then} $\bm{R}[i,:] \leftarrow \text{Dup}_b(\bm{M}[i,:] \bullet \bm{1}) \cdot \bm{\mu}$
	    \STATE \textbf{if} \textit{Case 2} \textbf{then} $\bm{R}[i,:] \leftarrow \text{Dup}_b(\bm{M}[i,:] \bullet \bm{\mu})$
	\ENDFOR
	\sbline
	\vspace{-1.5em}
	\end{algorithmic}
\end{method}

\subsubsection{\df{Further Optimizations}}
We note that all multiplication methods are parallelizable at the row level. Each multiplication of a ciphertext is followed by a rescale and a relinearization operation (the result of a multiplication with a plaintext only needs to be rescaled, see \S.\ref{subsec:mhe}). When the results of several multiplications need to be aggregated, we defer the rescale and relinearization operations until \emph{after} the aggregation step so they can be executed once overall, rather than for every multiplication. 
Because these operations account for between 52\% and 75\% of the multiplication time (Tab.~\ref{tab:localruntimes}) and \sys heavily relies on matrix multiplications, this optimization considerably improves \sys's overall performance. For example, it reduces the runtime of multiplying an encrypted $\bm{M}^{(8 \times 2^8)}$ with a cleartext $\bm{\tilde{N}}^{(2^8 \times 2^8)}$ with M\ref{alg:m1} from 24.6 to 3.8 seconds; the improvement is expected to be greater for larger matrices. 

\subsection{Matrix Transformations and Factorizations} \label{subsubsec:matrixtransfo}
\df{We introduce new routines for executing} sophisticated linear algebra operations required by the PCA on encrypted matrices and vectors.
We begin with the Householder transformation \cite{householder1958unitary}, a key building block in other matrix transformations such as QR$^T$ and Eigen, which we subsequently describe. \df{We also present a new algorithm, DQR$^T$, for executing a QR factorization on a matrix that is distributed among multiple parties.} Note that all methods except \df{DQR$^T$} require communication only for bootstrapping (DBootstrap; \S.\ref{subsec:mhe}), which has a negligible computation cost. The reported communication costs are thus measured by our optimized number of invocations of bootstrapping on a single ciphertext.

\begin{algorithm}[t]
\setcounter{algorithm}{1}
\footnotesize
	\caption{\textbf{- Encrypted Householder Vector (HH)}} 
	\label{alg:hh}
	\renewcommand{\thealgorithm}{}
	\begin{algorithmic}[1]
	\item[\textbf{Input:} Encrypted $\bm{v}^{(h \times 1)}$]
	\item[\textbf{Output:} Encrypted $\bm{v'}^{(h \times 1)}$, such that $\bm{H}=\bm{\tilde{I}}^{(h\times h)}-2\bm{v'} \times \bm{v'}^T$]
	\item[ensures $\bm{H}\times \bm{v}$ all zeros except the first coordinate]
	\item[\textbf{Comp. Cost:} $3 \cdot l(d) + 6 $ Mults and  $2 \cdot \log_2{(h)} $ Rots; $l(d)$ defined in text]
	\item[\textbf{Comm. Cost:} $(\frac{5+3(\ceil{1+\log_2(d)})}{\lambda}) \cdot \ceil{\frac{h}{t}}\ \text{Ciphertexts}$]
	\end{algorithmic}
	\vspace{-1.75em}
	\begin{multicols}{2}
	\begin{algorithmic}[1]
	\STATE $\bm{v^2} \leftarrow \bm{v} \cdot \bm{v}$ 
	\STATE $||\bm{v}||^2 \leftarrow \bm{v^2} \bullet 1$ 
	\STATE $||\bm{v}|| \leftarrow \sqrt{||\bm{v}||^2}$ 
	\vspace{0.25em}
	\STATE $\delta \leftarrow \bm{v}[0] / \sqrt{\bm{v}[0]^2} $
	\STATE $\delta \leftarrow \delta \cdot ||\bm{v}||$ 
	\STATE $\bm{u} \leftarrow \bm{v} $
	\STATE $\bm{u}[0] \leftarrow \delta + \bm{v}[0] $ 
	\STATE $\bm{u}^2 \leftarrow \bm{u} \cdot \bm{u}$ 
	\STATE $\bm{k} \leftarrow \bm{u}^2[0] + (||\bm{v}||^2- \bm{v^2}[0])$
	\STATE $\bm{k'} \leftarrow \text{Dup}_h(1 / \sqrt{\bm{k}[0]})$
	\STATE $\bm{v'} \leftarrow \bm{u} \cdot \bm{k'}$ 
	\vspace{-1.0em}
	\end{algorithmic}
	\end{multicols}
	\vspace{-1.0em}
\end{algorithm}

\noindent \textbf{Householder Transformation of Encrypted Vectors.}
Alg.~\ref{alg:hh} performs a key step in the Householder transformation, which reflects a vector about a given hyperplane, on an encrypted vector.
For use in PCA, we need to choose a specific reflection hyperplane that transforms the input vector $\bm{v}$ into a vector (of the same norm) with zeros in all coordinates except for the first.
The output $\bm{v'}$ of Alg.~\ref{alg:hh} represents this hyperplane; the Householder matrix obtained as $\bm{H}=\bm{\tilde{I}}^{(h\times h)}-2\bm{v'} \times \bm{v'}^T$, where $\bm{\tilde{I}}$ is the identity matrix, satisfies that $\bm{H}\times \bm{v}$ has a nonzero element only in the first coordinate.
This method is used in QR$^T$ to iteratively apply orthogonal transformations to the input matrix to convert it into a lower triangular matrix.
Following the standard technique, the norm of the input vector (computed in Lines~1-3) is added to or subtracted from its first coordinate (Line~7), depending on the sign of the first coordinate (Line~4) for numerical stability.
Afterwards, the vector is normalized (Lines 9-11) to obtain the desired reflection vector.

Alg.~\ref{alg:hh} requires the evaluation of non-polynomial functions, including the sign function (alternatively, $g(x)$\,=\,$x/\sqrt{x^2}$, Line 4), the square root, and the inverse square root. To this end, \sys applies Chebyshev polynomial approximation \cite{chebyshev1853theorie} to each function on a pre-determined input range (agreed upon in Step 1; \S.\ref{subsec:workflow}). In addition, we use the baby-step giant-step technique~\cite{Han_BetterBootstrap} to further reduce the complexity of evaluating degree-$d$ polynomials, resulting in a multiplicative depth of $\ceil{\log{(d)}+1}$ and $2 \cdot \sqrt{2d}+\frac{1}{2} \log_2{(d)}+\mathcal{O}(1)$ ciphertext multiplications. We denote this quantity as $l(d)$ in our algorithms. We discuss the choice of approximation intervals in \S.\ref{subsubsec:approx}.
For the communication cost, we calculate the number of DBootstrap executions as the multiplicative depth of this method divided by the number of available ciphertext levels $\lambda$ (\S.\ref{subsec:mhe}).

\noindent\textbf{QR Factorization of Encrypted Matrices.}
QR factorization decomposes an input matrix $\bm{V}$ into an orthogonal matrix $\bm{Q}$ and a lower-triangular matrix $\bm{R}$ such that $\bm{V} = \bm{R} \times \bm{Q}$.
This is repeatedly used in Steps 4, 6 and 7 of \sys's workflow.
In Alg.~\ref{alg:qrt}, we describe both the transposed-QR factorization QR$^T$ that is executed by one DP on an encrypted matrix and its distributed equivalent DQR$^T$. \df{DQR$^T$ performs a QR factorization in a federated manner on an encrypted matrix that is distributed among the DPs, requiring the DPs to aggregate (denoted by $\Xi$) their partial results in Lines~3, 5, and 18.}
In Step~4 of \sys, QR$^T$ is executed on a matrix with $h=m$ columns (i.e., same as the number of features), whereas DQR$^T$ is executed on a matrix with $h=n$ columns distributed among the DPs, where each DP$_i$ has $n_i$ columns. $\text{HH}$ and the vector-matrix multiplications in Lines 5 and 18 are the only operations with a cost that depends on $h$. DQR$^T$ requires more communication among the parties, and the complexity of QR factorization depends mainly on the number of rows $\delta$, which is the same in both QR$^T$ and DQR$^T$, and not on $h$. Hence, we use DQR$^T$ only when the difference between $n_i$ and $m$ is large enough to compensate for the communication overhead, i.e., when $\xi \cdot \log_2{(n_{\max})} < \log_2{(m)}$, with a factor $\xi$ determined by the properties of the network setup (e.g., latency). Note that $n_{\max} = \max_i (n_i)$.
\begin{algorithm}[t]
\footnotesize
	\caption{\textbf{ - Encrypted QR$^T$ Factorization (or DQR$^T$)}:}
	\label{alg:qrt}
	\renewcommand{\thealgorithm}{}
	\begin{algorithmic}[1]
	\item[\textbf{Input:} Encrypted $\bm{V}^{(\delta \times h)}$]
	\item[\textbf{Output:} Encrypted $\bm{Q}^{(\delta \times h)}\ \text{and}\ \bm{R}^{(\delta \times \delta)}$, such that $\bm{R}\times \bm{Q} = \bm{V}$] 
	\item[\textbf{Comp. Cost:} $\mathcal{O}(\delta^2 + \delta \cdot \zeta_{\text{HH}})\ \text{Mults}\text{,}\ \mathcal{O}(\delta^2 \cdot (1+\log_2(h)) + \delta \cdot \zeta_{\text{HH}})\ \text{Rots}$,]
	\item[where $\zeta_{\text{HH}}$ refers to the cost of HH (Alg.~\ref{alg:hh}).]
	\item[\textbf{Comm. Cost:} $C = \delta \cdot \zeta_{\text{HH}} + \frac{4\delta^2}{\lambda} \cdot \ceil{\frac{h}{t}}$ Cipher. (\textbf{DQR}: $C  + 3 \cdot \delta \cdot \ceil{\frac{h}{t}}$ Cipher.)]
    \end{algorithmic}
    \vspace{-1.5em}
    \begin{multicols}{2}
    \begin{algorithmic}[1]
    \STATE $\bm{H} \leftarrow \bm{0}^{(\delta \times h)}$
    \STATE \textbf{for} $i = 0,\ \dots,\ \delta-1$ \textbf{do}
    \STATE \Hquad $\bm{v} \leftarrow \text{HH}(\bm{V}[0,:]^T)$\\
	    \Hquad (\textbf{DQR}: $\Xi$ in Line 2 of Alg.~\ref{alg:hh})
        \STATE \Hquad $\bm{H}[i,:] \leftarrow \bm{v}^T$
        \STATE \Hquad $\bm{v'} \leftarrow \bm{V} \times \bm{v}$  (\textbf{DQR}: $\Xi( \bm{v'})$)
        \STATE \Hquad \textbf{for} $j = 0,\ \dots,\ \delta-i-1$ \textbf{do}
	        \STATE \quad $\bm{V}[j,:] \leftarrow \bm{V}[j,:]$\\ \quad $- 2 \cdot (\bm{v}^T \cdot \text{Dup}_{(\delta-i)}(\bm{v'}[j]))$
         \STATE \Hquad \textbf{end for}
	    \STATE \Hquad $\bm{r} \leftarrow \text{Rot}_{(-i)}(\bm{V}[0,:])$
	    \STATE \Hquad $\bm{R}[i,:] \leftarrow \bm{r}[:\delta]$
            \STATE \Hquad \textbf{for} $j =\ 0,\ \dots,\ \delta-i-1$ \textbf{do}
	        \STATE \quad $\bm{V}[j,:] \leftarrow \text{Rot}_{1}(\bm{V}[j+1,:])$
            \STATE \Hquad \textbf{end for}
        \STATE \textbf{end for}
	\STATE $\bm{Q} \leftarrow [ \bm{\tilde{I}}^{(\delta\times \delta)} \  \textbf{0}^{(\delta \times (h-\delta))} ]$
	\STATE \textbf{for} $i = \delta-1,\ \dots, 0$ \textbf{do}
	    \STATE \Hquad $\bm{H}[i,:] \leftarrow \text{Rot}_{(-i)}(\bm{H}[i,:])$ 
            \STATE \Hquad $\bm{h'} \leftarrow \bm{Q} \times \bm{H}[i,:]^T$ 
	\item[ \hspace{1.8em} (\textbf{DQR}: $\Xi(\bm{h'})$)]
        \STATE \Hquad \textbf{for} $j = 0,\ \dots,\ \delta-1$ \textbf{do}
	        \STATE \quad $\bm{Q}[j,:] \leftarrow \bm{Q}[j,:]$ \\ \quad $- 2 \cdot \bm{H}[i,:]^T \cdot \text{Dup}_{i}(\bm{h'}[j])$
        \STATE \Hquad \textbf{end for}
    \STATE \textbf{end for}
    \end{algorithmic}
    \end{multicols}
    \vspace{-1.5em}
\end{algorithm}

 From Lines~1 to 14, the input matrix $\bm{V}$ is multiplied by the Householder matrix $\bm{H} = \bm{\tilde{I}} - 2 \bm{v}\times \bm{v}^T$, where $\bm{v}=\text{HH}(\bm{V}[0,:]^T)$ is the Householder vector obtained by Alg.~\ref{alg:hh} with the first row of $\bm{V}$ as input.
 This transformation is recursively performed on the $(i,i)$ minors of $\bm{V}$ by discarding the first row and the first column to incrementally obtain the lower-triangular matrix $\bm{R}$.
 Due to \sys's vectorized encoding scheme, the sub-matrix is efficiently obtained by applying a single ciphertext rotation per row (Line 9).
 In \sys, $\bm{R}$ is only used during the eigendecompostion in Step 6.
 $\bm{Q}$ is computed in the second part (Lines 15 to 22) and corresponds to the product of all Householder matrices $\bm{H}$.
 
 Recall that we minimize bootstrapping by refreshing only small-dimensional data that are globally shared among the DPs (\S.\ref{sec:design}). The intermediate values in QR$^T$ satisfy this condition as they are derived from the input matrix that is already aggregated.
 Hence, the optimized number of invocations of DBootstrap($\cdot$) for QR$^T$ corresponds to its multiplicative depth divided by $\lambda$.
 For DQR$^T$, the input matrix is split among the DPs.
 In this case, the results of the collective aggregation ($\Xi$) in Lines~5 and 18, which constitute globally shared vectors among the DPs, are bootstrapped before being broadcast (shown as the additional cost).

\noindent\textbf{Eigendecomposition of Encrypted Matrices.}
Alg.~\ref{alg:eigendecompo} decomposes an encrypted matrix $\bm{M}$ into $\bm{Q}\times\bm{L}\times\bm{Q}^T$, where $\bm{Q}$ is a matrix of eigenvectors and $\bm{L}$ is a diagonal matrix with the diagonal defined by the encrypted vector of eigenvalues $\bm{l}$. The eigenvalues are ordered from the largest to the smallest. We adapt the standard QR iteration algorithm~\cite{wang2001convergence, ortega1963ll} to the setting with an encrypted input matrix. 
The encrypted matrix is first tridiagonalized, i.e., transformed to a matrix where the only nonzero elements are in the diagonal, the subdiagonal, or the superdiagonal, which is known to improve the convergence rate of eigendecomposition~\cite{wang2001convergence}.
The tridiagonalization is achieved by applying Householder transformations (using Alg.~\ref{alg:hh}) to different subparts of the matrix to introduce zeros (Lines~2 to 11 in Alg.~\ref{alg:eigendecompo}).
The resulting encrypted matrix $\bm{T}$ is then iteratively factorized using QR$^T$ (Line 17) into $\bm{R}\times \bm{Q'}$ (note the row-wise application of QR) and reconstructed as $\bm{Q'} \times \bm{R}$ to gradually transform the matrix into a diagonal matrix. 
During this process, the last diagonal element converges to the smallest eigenvalue of the input.
This is then executed for each eigenvalue in an ascending order, and the corresponding eigenvectors are obtained from the product of all $\bm{Q'}$ matrices.
We perform all small-matrix multiplications (Lines~6, 7, 8, 18) by encoding each matrix in a single ciphertext \df{and employing the technique of Jiang et al.~\cite{jiang2018secure}.  We refer to this method as M5 to distinguish from the large-scale, unbalanced setting in M3 with ciphertext-cleartext multiplications. 
Multiplying two $s \times s$ encrypted matrices requires $5s$ Mults and $3s + 5\sqrt{s}$ Rots.} We convert the matrices to our row-wise encoding scheme (in Lines 6, 9, 10, 12, 15, 17, 20, 22, 23) using one multiplication and one rotation per row, only to efficiently perform row and column selections. Similarly as Alg.~\ref{alg:qrt}, this method operates on globally shared inputs, and its optimized communication cost scales with the multiplicative depth divided by $\lambda$, in addition to the costs of the HH and QR$^T$ subroutines.

\begin{algorithm}[t]
\footnotesize
\caption{\textbf{- Encrypted Eigendecomposition (Eigen):}}
	\label{alg:eigendecompo}
	\renewcommand{\thealgorithm}{}
	\begin{algorithmic}[1]
	\item[\textbf{Input:} Encrypted symmetric $\bm{M}^{(\eta \times \eta)}$, number of iterations $w$]
	\item[\textbf{Output:} Encrypted $\bm{Q}^{(\eta \times \eta)}$ and $\bm{l}^{(1 \times \eta)}$, where the rows of $\bm{Q}$ are ]
	\item[eigenvectors of $\bm{M}$, and $\bm{l}$ has corresponding eigenvalues]
	\item[\textbf{Comp. Cost:} $\mathcal{O}(\eta \cdot (1 + \zeta_{\text{HH}} + w \cdot \zeta_{\text{QR}}) + \eta \cdot (1 + w \cdot \zeta_{\text{M5}})) \ \text{Mults}$ ]
	\item[and $\mathcal{O}(\eta \cdot (\zeta_{\text{HH}} + w \cdot \zeta_{\text{QR}}) + \eta \cdot(1 + (w \cdot \zeta_{\text{M5}}))) \ \text{Rots}$,]
	\item[where $\zeta_{\text{HH}}$, $\zeta_{\text{M5}}$ and $\zeta_{\text{QR}}$ refer to the costs of HH, M5 and QR$^T$.]
	\item[\textbf{Comm. Cost:} $(\eta-1) \cdot (\zeta_{\text{HH}} + w \cdot \zeta_{\text{QR}}) + \frac{(\eta-1)\cdot(4+3w)}{\lambda} \cdot \ceil{\frac{\eta}{t}}$ Ciphertexts]
	\end{algorithmic}
	\vspace{-1.0em}
	\begin{multicols}{2}
	\begin{algorithmic}[1]
    \STATE $\bm{Q} \leftarrow \bm{\tilde{I}}^{(\eta\times \eta)},\ \bm{T} \leftarrow \textbf{0}^{(\eta \times \eta)}$
    \FOR{$i= 0,\ \dots, \eta-3$}
    \STATE $\bm{v} \leftarrow \text{HH}(\bm{M}[0,1\text{:}]^T)$
    \STATE $\bm{P} \leftarrow \bm{\tilde{I}}^{((\eta-i) \times (\eta-i))}$
    \STATE $\bm{P}[1\text{:},1\text{:}] \leftarrow \bm{\tilde{I}}^{((\eta-i-1)\times (\eta-i-1))}$\\ \qquad \qquad \ \  $- 2 \cdot \bm{v} \times \bm{v}^T$
    \STATE $\bm{Q}[i\text{:},\text{:}]  \leftarrow \bm{P} \times \bm{Q}[i\text{:},\text{:}]$
    \STATE $\bm{PM}  \leftarrow \bm{P} \times \bm{M}$
    \STATE $\bm{M}  \leftarrow \bm{PM} \times \bm{P}^T$
    \STATE $\bm{T}[i\text{:}i+2,i\text{:}i+2]$$  \leftarrow$$\bm{M}[\text{:}2,\text{:}2]$
    \STATE $\bm{M} \leftarrow \bm{M}[1\text{:},1\text{:}]$
    \ENDFOR
    \STATE $\bm{T}[\eta-2\text{:},\eta-2\text{:}] \leftarrow \bm{M}$  
    \FOR{$i = \eta-1,\ \dots, 1$}
    \FOR{$j = 0, ..., w-1$}
    \STATE \hspace{-0.8em} $\bm{S} \leftarrow \bm{T}[i,i] \times \bm{\tilde{I}}[i,\text{:}]$
    \STATE \hspace{-0.8em} $\bm{T} \leftarrow \bm{T} - \bm{S}$
    \STATE \hspace{-0.8em} $\bm{Q'},\bm{R} \leftarrow \text{QR}^T(\bm{T})$
    \STATE \hspace{-0.8em} $\bm{T} \leftarrow \bm{Q'} \times \bm{R}$
    \STATE \hspace{-0.8em} $\bm{T} \leftarrow \bm{T} + \bm{S}$
    \STATE \hspace{-0.8em} $\bm{Q}[\text{:$i$+1,:}] \leftarrow \bm{Q}[\text{:$i$+1,:}] \times \bm{Q'}$
    \ENDFOR
    \STATE $\bm{l}[i] \leftarrow \bm{T}[i,i]$
    \STATE $\bm{T} \leftarrow \bm{T}[\text{:}i,\text{:}i]$
    \ENDFOR
    \STATE $\bm{l}[0] \leftarrow \bm{T}[0,0]$ \\ \ 
	\end{algorithmic}
	\end{multicols}
	\vspace{-2.0em}
\end{algorithm}

\subsection{Non-Polynomial Functions on Encrypted Inputs} \label{subsubsec:approx}
To approximate non-polynomial functions on chosen intervals, \sys's default approach is to rely on homomorphic evaluations of Chebyshev polynomial approximations \cite{Han_BetterBootstrap}. In Step~1 (Alg.~\ref{alg:workflow}), the DPs agree on the intervals and on the degree of the approximations. The complexity of the polynomial evaluation increases with the degree but is independent of the interval size, which influences the precision. While any interval selection approach may be used with \sys, the approach we adopt in our evaluation in \S.\ref{sec:eval} is for a DP (e.g., the one coordinating the collaboration or the one with the highest number of local samples) to set the intervals based on the estimated range of intermediate values to be encountered by running RPCA on a simulated dataset, obtained by upsampling its local data to match the size of the joint data. 
In \S.\ref{sec:extensions}, we discuss an extension to \sys that enables it to switch to secret sharing for the evaluation of non-polynomial functions, for which efficient bit-wise protocols exist for scaling the input to a common range for approximation.
This effectively removes the need to choose intervals and, depending on the parameters, can further improve \sys's accuracy (Appendix \ref{app:mheLss}).
\section{System Evaluation}\label{sec:eval}
We show that \sys, enabled by our optimization techniques~(\S.\ref{sec:design}), efficiently computes a PCA on high-dimensional inputs distributed among a large number of DPs.
We demonstrate \sys's practicality and accuracy on various datasets with the number of features ranging from 8 to 23,724 and including up to 60,000 samples. \sys consistently obtains PCs that are highly similar ($r^2>0.9$) to those obtained by a standard non-secure PCA. \sys also outperforms alternative privacy-preserving approaches in terms of accuracy and runtime, and offers stronger security guarantees compared to some. In \S.\ref{subsec:evalAppl}, we show that, contrarily to meta-analysis, \sys remains accurate regardless of potential differences in the data distribution among the DPs.

\subsection{Formal Analysis of Costs}\label{subsec:theo}
\sys's communication cost depends mainly on the number of features $m$, the number of components $\rho$ and the number of power iterations $p$. \sys's computation cost depends on the same parameters and optionally on the number of samples per DP $n_i$. For both, the overall cost is amortized over the ciphertexts due to packing and the SIMD property of HE, effectively dividing the contributions of $m$ and $n_i$ to the complexity by the ciphertext capacity $t$. In Tab.~\ref{tab:complexity}, we show the theoretical costs for a single DP (DP$_i$) for each step in \sys (Alg.~\ref{alg:workflow}).

\df{The communication in Step 1 is due to the generation of the public key $pk$ and evaluation keys $evks$ (including a relinearization key and $\log_2(t)$ rotation keys). All rotations in \sys are executed by combining rotations of power-of-two shifts using the pre-generated keys.} $\text{DBootstrap}$ requires each DP to transmit and receive \df{the equivalent of a} ciphertext, and to perform one ciphertext addition (at a negligible cost). \df{In the remaining steps, } we analyze the communication cost in terms of the number of $\text{DBootstrap}$ invocations, which depends on the cryptographic parameters and the number of multiplications to perform in each routine (\S.\ref{subsubsec:matrixtransfo}).
In turn, the number of multiplications depends on the input dimensions, the degree of polynomial approximations and, for $\text{Eigen}$, the number of iterations $w$.

\sys's overall communication cost is independent of the number of samples and is dominated by the bootstrapping execution. We optimize the performance of \sys by selecting the computation approach with the lowest complexity, e.g., by choosing $\textsl{Precomp}$ (whose complexity is independent of $n_i$) if the number of samples is large. 

\begin{table}[t]
\center
\scriptsize
\setlength\tabcolsep{1.5pt}
\begin{tabular}{ |c|c|c| } 
 \hline
   \textbf{Step} & \textbf{Comm.}  & \textbf{Computation}  \\
 \toprule \hline
 1 & $\log_2(t)+2.5$ & -  \\
 \hline
 2 & $1 \cdot \ceil{\frac{m}{t}}$ & -  \\
 \hline
 3 & $\rho \cdot \ceil{\frac{m}{t}}$ & -  \\
 \hline
 4 & $p \cdot (\rho \cdot \ceil{\frac{m}{t}}  $ & $p \cdot ((\textsl{Precomp}\text{ or }\textsl{Seq}) + \zeta_{\text{QR}}(\rho, m))$; $\textsl{Precomp} = \zeta^*_{\text{M}}(\rho,m, m)$     \\
  & $+ \zeta_{\text{QR}}(\rho, m))$ & $\textsl{Seq} = \zeta^*_{\text{M}}(\rho,m, n_i) + \zeta^*_{\text{M}}(\rho,n_i, m)$ \\
 \hline
 \multirow{2}{*}{5} & \multirow{2}{*}{$\rho \cdot \ceil{\frac{\rho}{t}}$} &  $\textsl{Precomp}$: $\zeta^*_{\text{M}}(\rho,m, m) + \zeta^*_{\text{M}}(\rho,m, \rho)$ \\
  & & $\textsl{Seq}$: $\zeta^*_{\text{M}}(\rho,m, n_i) +\zeta^*_{\text{M}}(\rho,n_i, \rho)$ \\
 \hline
 6 & $\zeta_{\text{Eigen}}(\rho, \rho)$ & $\zeta_{\text{Eigen}}(\rho,\rho)$ \\
 \hline
 7 & $\psi \cdot \ceil{\frac{m}{t}} $ & \textsl{Precomp}: $\zeta^*_{\text{M}}(\psi,\rho, m) + \zeta_{\text{QR}}(\psi,m)$ ; \textsl{Seq}:  \\
  & $+ \zeta_{\text{QR}}(\psi, m)$ & $\zeta^*_{\text{M}}(\psi,\rho, n_i) + \zeta^*_{\text{M}}(\psi, n_i, m) + \zeta_{\text{QR}}(\psi, m)$ \\
 \hline
 8 & - & $\zeta^*_{\text{M}}(\psi,m, n_i)$ \\
 \hline
\end{tabular}
\vspace{-0.5em}
\caption{\textbf{Communication and computation costs of \sys (Alg.~\ref{alg:workflow}).} $\zeta_{x}(\text{dim})$ returns the cost of the function $x$ according to the dimensions (dim). The functions' costs are defined in \S.\ref{subsub:matmult} for M, in Alg.~\ref{alg:qrt} for QR and in Alg.~\ref{alg:eigendecompo} for Eigen.}
\label{tab:complexity}
\end{table}

\subsection{Implementation Details \df{and Evaluation Settings}}\label{subsec:implementation}
We implemented \sys in Go~\cite{Go}, building upon Lattigo~\cite{lattigo} and Onet~\cite{Onet}, which are open-source Go libraries for lattice-based cryptography and decentralized system development, respectively. The communication between DPs is through secure TCP channels (using TLS). We evaluate our prototype based on a realistic network emulated using Mininet~\cite{mininet}, with a bandwidth of 1~Gbps and a communication delay of 20ms between every two nodes. \df{Unless otherwise stated, we uniformly and horizontally distribute the input data among 6 DPs. We deploy each DP on a separate Linux machine with Intel Xeon E5-2680 v3 CPUs running at 2.5~GHz with 24 threads on 12 cores and 256 GB of RAM.}
\df{We provide the default system parameters of \sys considered in our evaluation in Appendix \ref{app:symbols}}. 
 
\subsection{Microbenchmarks for MHE Protocols in \sys 
} \label{subsec:microbench}
In Tab.~\ref{tab:localruntimes}, we summarize the runtimes for \sys's main ciphertext operations as well as high-level linear algebra routines. Recall that each ciphertext contains up to $t=2^{13}$ values and any operation is concurrently executed on all encrypted values. Multiplying a cleartext with a ciphertext is almost 8x faster than multiplying two ciphertexts with the default parameters. 
The transmission time of a ciphertext (Send($c$)) depends mostly on the communication delay (20ms in our setting). \df{In our default setting, $\text{DKeyGen}$ takes 9 seconds to generate the public key, relinearization key, and 13 rotations keys.}

\subsection{Practical Scalability of \sys Performance} \label{subsec:scalability}
We evaluated \sys's scalability on simulated datasets of varying sizes.
In Fig.~\ref{fig:perffeaturesSamples}, we show that \sys's runtime (when computing eight PCs with ten power iterations) remains almost constant when the dimensions are smaller than the ciphertext capacity $t$ (set to 8,192 by default). It then grows linearly with the number of ciphertexts, i.e., with the number of features ($m$) and samples per DP ($n_i$) divided by $t$. \df{Note that, since the protocol is synced among the DPs  at each aggregation step, \sys's runtime depends on the slowest DP, e.g., the DP with the largest local dataset, as shown in Fig.~\ref{fig:perffeaturesSamples} and Fig.~\ref{fig:maxNbrOfSamples} in appendix.} In all figures, we omit the negligible execution times of Steps 1 to 3. These steps require mostly non-iterative cleartext operations. In Fig.~\ref{fig:perffeaturesSamples} (left panel), we set  $n_i = 1,024$ and show that all \sys's approaches (i.e., \textsl{Precomp} and \textsl{Seq} with QR$^T$ or DQR$^T$) similarly scale linearly with $m$. \textsl{Precomp} is the most efficient approach for this range of values for $m$ and $n=6,144$ but becomes impractical with a large $m$. In these experiments, we found DQR$^T$ to be consistently inferior to QR$^T$ as the former's communication overhead overshadows its computational speedup. This is expected, since the computational gain of using DQR$^T$ depends on how much smaller $n_i$ is with respect to $m$ (see \S.\ref{subsubsec:matrixtransfo}). This difference is never large enough to compensate for the communication overhead in our settings. The results in Fig.~\ref{fig:perffeaturesSamples} (right panel), for which $m$ is set to 256, show that \sys's runtime remains constant when using \textsl{Precomp}, which does not depend on the number of samples.

\begin{table}[t]
\scriptsize
\setlength\tabcolsep{1.5pt}
        \centering
        \begin{tabular}{|c|c||c|c|}
                \hline
                \textbf{Operation} & \textbf{Runtime (s)} & \textbf{Operation} & \textbf{Runtime (s)} \\ \toprule \hline
                $+$ & $7 \cdot 10^{-4}$ &  M\ref{alg:m2} - ($\bm{M}^{(8 \times 2^8)} \times \bm{\tilde{N}}^{(2^8 \times 2^8)}$) & $59$ \\ \hline
                $\tilde{\bm{v}} \cdot \bm{c}$ & $0.013$ & M\ref{alg:m3} - ($\bm{M}^{(8 \times 2^8)} \times \bm{\tilde{N}}^{(2^8 \times 2^8)}$)& $51$ \\ \hline
                $\bm{c} \cdot \bm{c}'$ & $0.083$ & M\ref{alg:m1} - ($\bm{M}^{(8 \times 2^8)} \times \bm{\tilde{N}}^{(2^8 \times 2^8)}$) & $3.8$ \\ \hline
                $\text{Rot}(\cdot)$ & $0.08$  & M\ref{alg:mv} ($\bm{M}^{(8 \times 2^8)} \times \bm{\mu}^{(1 \times 2^8)}$) & $0.7$ \\ \hline
                $\bm{c}^{(1 \times 2^8)}\bullet \bm{c}'^{(1 \times 2^8)}$ & $0.73$ & M5 - ($\bm{M}^{(8 \times 8)} \times \bm{\tilde{N}}^{(8 \times 8)}$)& $0.9$ \\ \hline
                QR$^T(\bm{M}^{(8 \times 2^8)})$ & $117$ & Send($\bm{c}$) & $0.026$ \\ \hline
                DQR$^T(\bm{M}^{(8 \times 2^8)})$   & $227$ & $ \text{DBootstrap}(\cdot)$ & $0.49$ \\ \hline
                QR$^T(\bm{M}^{(8 \times 8)})$   & $94$ & \df{DKeyGen$(\cdot)$} & \df{9.0} \\ \hline 
                \end{tabular}
                \vspace{-0.5em}
        \caption{\textbf{\sys's micro-benchmarks with default parameters.} Send($\bm{c}$) transmits a ciphertext $\bm{c}$ from one DP to another. \df{M5 refers to the small encrypted matrix multiplication in \S.\ref{subsubsec:matrixtransfo}.}}
        \label{tab:localruntimes}
\end{table}

We remark that \sys's runtime is dominated by the time dedicated to the communication between the DPs (Fig.~\ref{fig:perfPCsDPS}). The communication overhead ranges between 90\% of the runtime with small input dimensions, i.e., when the packing capacity of the cryptoscheme is less exploited, and 45\% of the runtime when the dimensions are equal or larger than $t$. Although \sys is able to minimize its computation runtime with optimized federated and parallelized computation methods, its communication overhead is bounded by the available communication network. When the number of DPs ($s$) doubles, \sys's runtime increases only by a factor of around 1.1. This is because the amount of local computation does not grow with $s$ (Table~\ref{tab:complexity}) and because the cost of interactive routines only slightly increases with $s$. 
Based on Fig.~\ref{fig:perfPCsDPS}, we estimate practical runtimes even for hundreds of DPs, e.g., 110 minutes for 200 DPs with a maximum of 1,024 data samples per DP. We discuss in \S.\ref{sec:extensions} how \sys can be extended to handle availability issues given many DPs. In Fig.~\ref{fig:perfPCsDPS}, \sys's runtime grows linearly with the number of components in all its steps except in Step 6, where the eigendecomposition cost depends on the small matrix dimensions: $\rho \times \rho$. \sys's runtime increases linearly with the number of power iterations; however, this parameter typically does not grow with the data size for RPCA.

\begin{figure}[t]
    \centering
        \includegraphics[width=1.0\columnwidth]{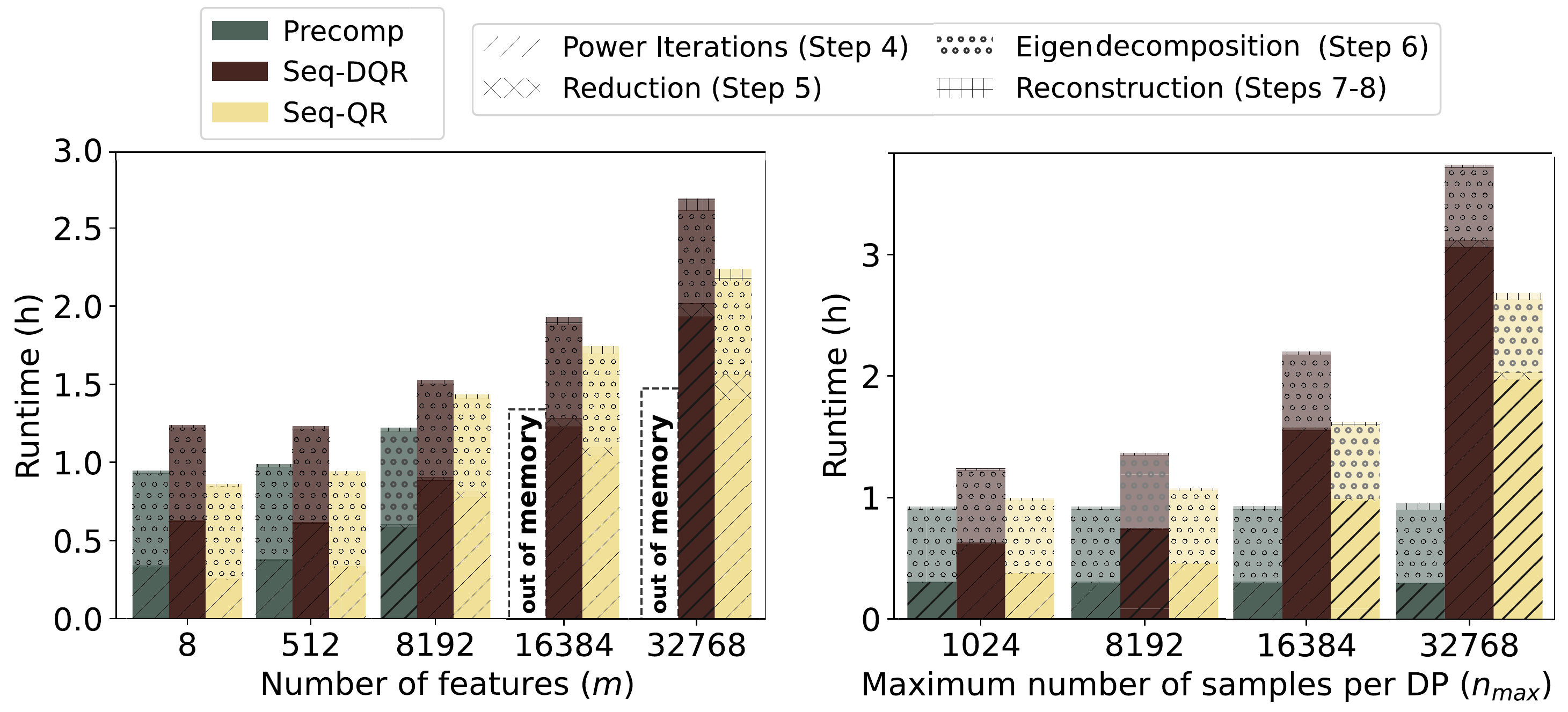}
        \vspace{-1.5em}
        \caption{\textbf{Runtime scaling with the number of features and samples.}}
        \label{fig:perffeaturesSamples}
\end{figure}
\begin{figure}[t]
    \centering
        \includegraphics[width=1.0\columnwidth]{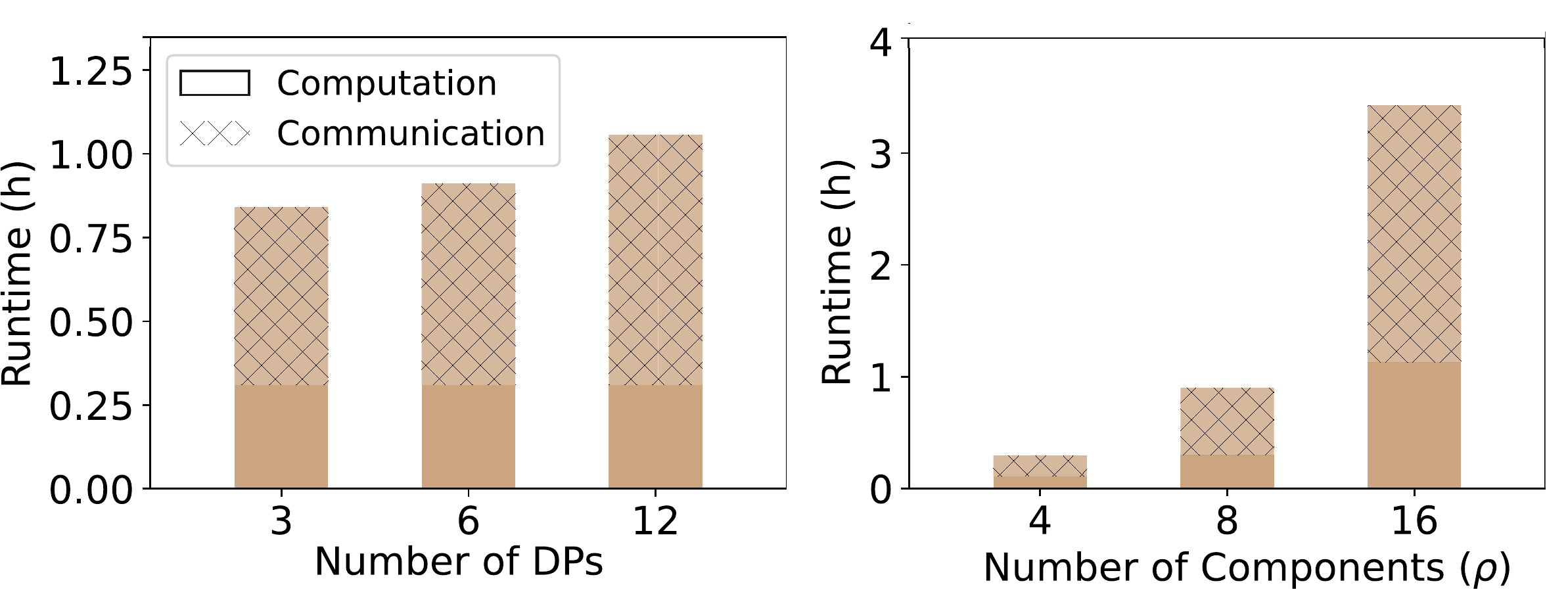}
        
        \vspace{-0.5em}
        \caption{\textbf{Runtime with the number of DPs and components.}}
        \label{fig:perfPCsDPS}
\end{figure}
In our default scenario, \sys's runtime is multiplied by a small factor of 1.1x when the available bandwidth is halved and the communication delay doubled. Moreover, each ciphertext accounts for 2.5~MB, thus executing \sys on a dataset with 8,192 features (or less) requires each DP to send 3.8~GB, independently of the number of samples $n$ (which can be large). 

\subsection{Accuracy of \df{SF-PCA} Results} \label{subsec:accuracy}
We demonstrate \sys's accuracy and practicality on six real datasets, including MNIST \cite{MNIST} and two genomic datasets \cite{lungcancer, 1000genomes} with thousands of patients and up to 23,724 features (see Appendix \ref{app:datasets} for dataset details). 
We evenly and randomly split each dataset among the DPs. In  \S.\ref{subsec:evalAppl}, we show that \sys computes the same results regardless of the data distribution among the DPs. 
In Tab.~\ref{tab:accuracy}, we show that \sys and the cleartext non-secure centralized Randomized PCA (RPCA, Fig.~\ref{fig:randoPCA}) achieve similar accuracy (according to the mean-squared error, MSE; and Pearson Correlation Coefficient, $r^2$), with respect to the PCs obtained using the standard non-secure PCA, i.e., the RPCA implemention provided by the \emph{sklearn} Python package~\cite{sklearn}.

\subsection{Comparison with Existing Works}\label{subsec:comparison}
We compare \sys with existing approaches for federated or multiparty PCA, which we categorize into meta-analysis, centralized HE (C-HE), and secret sharing-based SMC solutions. A more detailed review of these approaches is provided in \S.\ref{sec:related}.

\noindent\textbf{Meta-analysis.} For comparison, we replicate the meta-analysis approach of Liang et al \cite{liang2014improved}, whereby a central computing server performs a truncated SVD on the combined SVD results obtained independently by each DP. 
In Tab.~\ref{tab:accuracy}, we show that this solution yields the least accurate results across all datasets. 
\df{Note that \sys significantly improves upon the accuracy of meta-analysis by emulating a centralized PCA.}
Moreover, most meta-analysis solutions \cite{abu2002distributed, bai2005principal,Balcan2014,Cheung, fan2019distributed, fellus2014dimensionality, gang2019fast, liang2013distributed,liang2014improved,qi2004global, Won2016} are not end-to-end secure as the DPs' intermediate results are revealed to an aggregator server (or to the other DPs). These solutions achieve similar runtimes as non-secure centralized solutions because they also operate on unprotected cleartext data.

\noindent\textbf{Centralized HE (C-HE).} We estimate the runtime of an HE-based centralized solution based on \sys's runtime as follows. We account for the fact that the computations cannot be distributed among the DPs and that all operations must be performed on the encrypted data. \df{Recall that \sys exploits local cleartext operations to optimize computation (e.g., \S.\ref{subsub:matmult}) and that multiplying two ciphertexts is 8 times slower than a plaintext-ciphertext multiplication.} We also include the overhead brought by a centralized bootstrapping routine \cite{Han_BetterBootstrap}, which is two orders of magnitude slower than $\text{DBootstrap}$, e.g., 26 seconds for \cite{Han_BetterBootstrap} vs. 0.49 seconds with $\text{DBootstrap}$. \df{Furthermore, since centralized bootstrapping consumes levels and lowers the number of available levels for multiplications, }C-HE would require more conservative cryptographic parameters with larger ciphertexts, and thus higher computation and communication costs. In Tab.~\ref{tab:accuracy}, we show the estimated lower bound of the runtime for a C-HE solution executed by a single DP. We remark that \sys, by distributing its workload and relying on efficient interactive protocols, is consistently 1-2 orders of magnitude faster than a C-HE solution. \df{We note that we consider C-HE solutions based on the same underlying scheme of \sys with comparable parameters, and that more sophisticated centralized solutions could be devised. However, those would still suffer from a high communication overhead and introduce a single point of failure due to the data centralization. 
}
\begin{figure*}[t]
\vspace{-2.0em}
\scriptsize
        \centering

\subfloat{
        \setlength\tabcolsep{0.8pt}
            \adjustbox{valign=b}{
            \begin{tabular}{ |c|c|c|c|c|c|c|c|c|c| }
            \hline
             & \multirow{3}{*}{\textbf{\parbox{1.0cm}{\centering End-to-end confidentiality}}}      & \multirow{3}{*}{\textbf{\parbox{1.1cm}{\centering No single-trusted DP}}} & \multirow{3}{*}{\textbf{\parbox{1.0cm}{\centering Locality of private data}}}      & \multicolumn{6}{|c|}{ \begin{tabular}[x]{@{}c@{}} \textbf{Dataset} ($n \times m$, $p, \psi, \alpha, w$) \\ MSE; $r^2$; Runtime (hours) \end{tabular}} \\ \cline{5-10}
             & & &  &  Wine (4,898$\times$11, & Lung (9,098$\times$23,724, & Pima (767$\times$8)\, & Chr20 (2,502$\times$1,773,  & MNIST (60,000$\times$760, & Vehicle (435$\times$19, \\
             & &  &  &  15, 5, 5, 1) \cite{wine} & 15, 2, 5, 5) \cite{lungcancer} & 5, 1, 1, 5) \cite{pima} & 20, 2, 4, 1) \cite{1000genomes} & 40, 5, 2, 5) \cite{MNIST} & 20, 8, 4, 5) \cite{vehicle} \\ \hline
            \textbf{RPCA} & \xmark & \xmark & \xmark & $<10^{-6};\ 0.99$; - & $0.05;\ 0.99$; - & $<10^{-11};\ 0.99$; - & $<10^{-6};\ 0.99$; - & $0.3;\ 0.99$; - & $0.15;\ 0.99$; - \\ \hline
            \textbf{Meta-an.} & \xmark & \xmark & \cmark & $<10^{-2};\ 0.99$; -  & $6.55;\ 0.7$; - & $<10^{-2};\ 0.98$; - & $0.32;\  0.98$; - & $1.48;\ 0.75$; - & $0.68;\ 0.99$; - \\ \hline
            \textbf{SF-PCA} & \cmark & \cmark & \cmark & $<10^{-6};\ 0.99$; 0.8 & $0.05;\ 0.99$; 3.5  & $<10^{-11};\ 0.99$; 0.04 & $<10^{-6};\ 0.99$; 0.79 &  $0.6;\ 0.91$; $2.22$ & $0.24;\ 0.99$; 0.77 \\ \hline
            \textbf{C-HE} & \cmark & \xmark & \xmark & -; -; $>11$ & -; -; $>46$ & -; -; $>0.7$ & -; -; $>12$ & -; -;  $>30$ & -; -; $>10$ \\ \hline
            \textbf{SMC} & \cmark & \xmark & \xmark & -; -; 3.0 & -; -; 9.6 & -; -; 0.39 & -; -; 2.32 & -; -; $22.66$ & -; -; $6.67$\\ \hline
\end{tabular}}\vspace{-1.0em} \label{tab:accuracyNew}}
\vspace{-0.5em}
        \caption{\textbf{Comparison with existing works on six real datasets.} MSE: mean-squared error, $r^2$:  Pearson correlation coefficient compared with ground truth PCs.}
        \label{tab:accuracy}
      \vspace{-1.5em}
\end{figure*}
\noindent\textbf{Secret Sharing-based SMC.}
In Tab.~\ref{tab:accuracy}, we compare \sys's runtime with the linear (additive) secret sharing-based SMC solution proposed by Cho et al. \cite{cho2018secure}. In this solution, two computing servers perform PCA on secret-shared data, and a third server is responsible for the generation and distribution of correlated random numbers used in SMC protocols (e.g., Beaver triples \cite{beaver1991efficient}). This additional party is trusted to correctly generate these values and not to collude with any other party. We ran Cho et al.'s publicly available, two-party solution~\cite{pcaHoon} in our evaluation environment. We further estimated the runtime of this solution with 6 DPs under linear scaling with the number of DPs.
We observe that \sys is between 3x and 10x faster than the SMC solution while operating in a stronger threat model without the need for an honest third party. We also note that the SMC solution requires the entire dataset to be secret-shared among the computing parties, which can be costly for large datasets and complicate regulatory compliance. For example, with the Lung dataset \cite{lungcancer}, this represents a communication overhead of more than 60~GB. 
Finally, we note that SMC solutions heavily rely on interactive computations, leading to many rounds of communication in total. Since a large portion of \sys is local non-interactive computation by each DP, \sys remains practical even in constrained networks with high communication delays, unlike the SMC solutions. For example, when we double the delay from 20ms to 40ms, we observed that \sys's runtime remains almost constant, whereas the SMC solution becomes 1.9 times slower in the two-party setting. In \S.\ref{sec:extensions}, we describe an extension of \sys which uses secret sharing specifically for non-polynomial operations over low-dimensional inputs. 

\subsection{Example Application of \df{SF-PCA} in Genomics}\label{subsec:evalAppl}
To further demonstrate the utility of \sys, we used it to analyze a genomic dataset of 2,504 individuals with 1,773 features (a subset of genetic variants from chromosome 20). PCA is a standard step in many genomic analysis workflows, e.g., in genome-wide association studies~\cite{price2006principal}, for capturing ancestry patterns in a dataset.
We split the data among three DPs such that each DP only has samples belonging to a specific ancestry group (Fig.~\ref{fig:chr20}.d).
The plots show individual samples projected onto the first two PCs. Consistent with the quantitative evaluation in \S.\ref{subsec:accuracy}, \sys (Fig.~\ref{fig:chr20}.b) is able to accurately identify the low-dimensional structure spanned by the data samples, almost exactly replicating the output of a centralized cleartext PCA on the full dataset, independently of how the data is split among the parties (Fig.~\ref{fig:chr20}.a). The meta-analysis approach for PCA (Fig.~\ref{fig:chr20}.c) results in a distorted data landscape due to the limited view of each DP.
In Fig.~\ref{fig:chr20}.d, we highlight the output of \sys that is visible to one of the DPs; while all DPs obtain projected data according to a \emph{unified} subspace identified by the PCA, each DP sees only a portion of the output associated with their local data as required by our security model.

\begin{figure}[h]
\vspace{-1.0em}
    \centering
    \small
    \includegraphics[width=1.0\columnwidth]{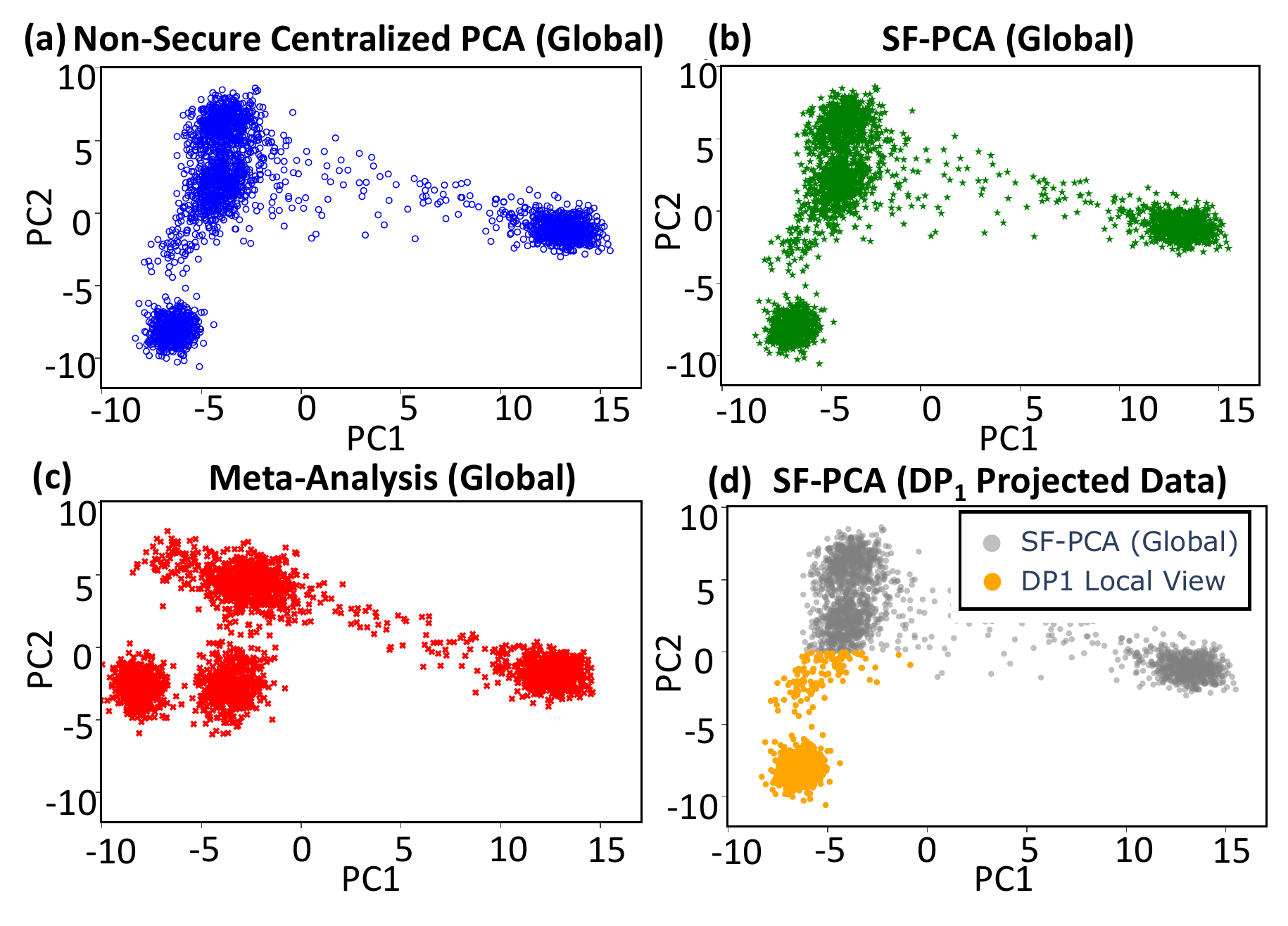}
    \vspace{-2.0em}
    \caption{{Demonstration of \sys on Genomic Data.}}
    \label{fig:chr20}
    \vspace{-1.5em}
\end{figure}
 \df{
 \section{Extensions}\label{sec:extensions}
 \sys can be extended in several ways to incorporate additional features. 
First, \sys's multiparty construction enables it to seamlessly and securely (i.e., without decryption) switch between MHE and secret sharing-based SMC  \cite{keller2018overdrive,bogdanov2008sharemind, shamir1979share} (see Appendix~\ref{app:mheLss}).
This enables \sys to leverage more efficient and accurate protocols to evaluate non-polynomial functions (e.g., sign tests) on small-dimensional inputs, while using MHE for operations over large encrypted vectors and matrices where the SIMD property of MHE leads to efficient performance with minimal communication.
 Next, the modular design of \sys enables its federated routines to be used to perform RPCA on vertically partitioned data (Appendix~\ref{app:verti}). \sys could also be extended to provide differential privacy (Appendix~\ref{app:diffP}), although setting a meaningful privacy parameter may be difficult, by incorporating an interactive protocol in which the DPs sequentially shuffle an encrypted list of noise values before adding them to the results upon decryption \cite{drynx}. Lastly, to cope with the possibility of a subset of DPs becoming unavailable during the PCA computation---particularly relevant for the setting with many DPs, \sys can be instantiated with a threshold secret sharing of the MHE secret key \df{\cite{mouchet2022efficient}} to allow a subset of DPs to continue the protocol execution (Appendix~\ref{app:fault}).
 }
\section{Discussion and Conclusions}\label{sec:conclusion}
We introduced \sys, a decentralized system for securely and efficiently executing PCA on data held by multiple data providers.
\sys ensures input confidentiality as long as at least one DP is honest.
Furthermore, the local private data never leave the DPs' premises given the federated design of \sys.
Our system builds on a range of optimized MHE-based routines we developed for key computational operations in PCA such as large-scale cleartext-ciphertext matrix multiplications and sophisticated linear algebra transformations, including matrix factorization and orthogonalization.
\sys obtains accurate results within practical runtimes on large matrices including tens of thousands of features, and efficiently scales with the number of data providers and the input dimensions due to our optimization strategies.

Our work shows that an end-to-end secure solution for high-complexity data analysis tasks such as PCA is practically feasible.
Incorporating \sys into existing privacy-preserving federated analysis methods (e.g., see Appendix~\ref{app:secureML}) and deploying it in a range of practical applications are natural next steps for our work.
Our design principles and optimization techniques that have led to the practical performance of \sys, as well as the optimized MHE routines for key linear algebra operations such as eigendecomposition are broadly applicable to other problems in federated analytics.

\section{Acknowledgements}
We thank Louis Vialar and the reviewers for their comments. This work was partially supported by NIH R01 HG010959 (to B.B.) and by NIH DP5 OD029574, RM1 HG011558, and Broad Institute's Schmidt Fellowship (to H.C.). J.R.T.-P. and J.-P.H. are co-founders of the start-up Tune Insight. All authors declare no other competing interests.

\let\OLDthebibliography\thebibliography
\renewcommand\thebibliography[1]{
  \OLDthebibliography{#1}
  \setlength{\parskip}{0pt}
  \setlength{\itemsep}{0pt plus 0.3ex}
}

\bibliographystyle{ieeetran}
\bibliography{IEEEabrv,bibfile_short}


\appendices

\renewcommand{\thesectiondis}[2]{\Alph{section}:}
\vspace{-0.5em}

\section{\df{CKKS}}\label{appendix:mhe}
\vspace{-0.5em}
We instantiate \sys's multiparty scheme with the Cheon-Kim-Kim-Song (\textsc{ckks}) cryptosystem~\cite{cheon2017homomorphic}. \textsc{ckks} parameters are denoted by the tuple $(\mathcal{N}, \Delta, \eta, mc)$, where $\mathcal{N}$ is the ring dimension; $\Delta$ is the plaintext scale by which any value is multiplied before it is quantized and encrypted/encoded; $\eta$ is the standard deviation of the noise distribution; and $mc$ represents a chain of moduli $\{q_0, \dots, q_L \}$ such that $\Pi_{\iota \in \{0, \dots, \kappa \}} q_\iota = Q_\kappa$ is the ciphertext modulus at level $\kappa$, with $Q_L$ the modulus of fresh ciphertexts. Operations on a ciphertext $\bm{c}$ at level $\kappa$ and scale $\Delta$ with $\Delta < Q_\kappa$ are performed modulo $Q_\kappa$. We denote by $\{\bm{c}, L, \Delta \}$, with $\bm{c}=(\bm{c}_{0}, \bm{c}_{1}) \in R^{2}_{Q_{L}}$, and $\bm{\tilde{p}}\in R_{Q_{L}}$, a fresh ciphertext at level $L$ with scale $\Delta$ and a plaintext, respectively.
\vspace{-0.5em}
\section{Symbols \& Default Values}\label{app:symbols}

\begin{table}[ht!]
\vspace{-1.0em}
\center
 \footnotesize
 \setlength\tabcolsep{1.0pt}
 \begin{tabular}{ |c|c|c| } 
  \hline
   \textbf{Symbol} & \textbf{Definition} & \textbf{Default}  \\
  \toprule
  \hline
   $s$, $p$, $w$  & \# DPs, \# power and eigen iterations & 6, 10, 5   \\
  \hline
  $\psi$ + $\alpha$ = $\rho$ & \# PCs + oversampling = \# components & 4 + 4 = 8   \\
  \hline
  $\zeta*$, $\zeta$, $d$ & Optimized cost, cost, approx. degree & -, -, 31   \\
  \hline
  $m, n, n_i$ & \# features, \# samples tot. \& at DP$_i$ & $2^8$, 6144, $2^{10}$   \\
  \hline
  $\mathcal{N}$, $\lambda$ & Ring dim., \# available levels & $2^{14}$, 7  \\ \hline 
  $R_Q$ & Plain/Ciphertext domain & - \\ \hline
 $\bm{c}$ & encrypted vector/ fresh ciphertext with & -\\
   & $\bm{c}=(\bm{c}_{0}, \bm{c}_{1}) \in R^{2}_{Q_{L}}$ 
   & \\ \hline
  $\bm{\tilde{p}}\in R_{Q_{L}}$, $sk,\ pk$ & plaintext, secret \& public keys  & -, -, -\\ \hline
  $t$, $\bullet$  & $\bm{c}$ capacity, dot product & $2^{13}$, -\\ \hline
  $\bm{M}^{(a \times b)}$, $\bm{\tilde{N}}^{(b \times c)}$  & Generic encrypted and cleartext matrices & -, - \\ \hline
 $\bm{M}[i,j]$, $\bm{v}[i]$  & Matrix/vector elem. at index $(i,j)$/$i$) & -, - \\ \hline
 \end{tabular}
 \vspace{-0.5em}
 \caption{\small{\textbf{Glossary of Symbols and Their Default Values in \sys.}}}
 \label{tab:symbols}
 \vspace{-0.5em}
 \end{table}
 
\vspace{-0.5em}
\section{\df{Security Analysis}}\label{sec:securityAnalysis}
 We rely on the real/ideal simulation paradigm~\cite{lindell2017simulate} to show that \sys achieves the input confidentiality requirement defined in \S.\ref{sec:overview}. A computationally bounded adversary that controls up to all but one DP cannot distinguish a \emph{real} world experiment, in which the adversary is given actual data from an execution of our protocol from the views of the compromised DP(s), and an \emph{ideal} world experiment, in which the adversary is given random data generated by a simulator.

The semantic security of the \textsc{ckks} scheme used in \sys is based on the hardness of the decisional-\textsc{rlwe} problem~\cite{cheon2017homomorphic,lyubashevsky2010ideal,Lindner2011}. Mouchet et al.~\cite{mouchet2019distributedbfv} proved that their distributed protocols, i.e., $\text{DKeyGen}$ and $\text{DKeySwitch}$, are secure under the simulator paradigm. They show that the distribution of the cryptoscheme preserves its security in the passive-adversary model with all-but-one dishonest DPs, as long as the decisional-\textsc{rlwe} problem is hard. Their proofs are based on the \textsc{bfv} cryptoscheme; Froelicher et al. \cite{spindle} showed that the proofs still hold with \textsc{ckks}, as the same computational assumptions hold, and the security of \textsc{ckks} is based on the same hard problem as \textsc{bfv}. They make a similar argument for $\text{DBootstrap}$ and prove its security.
The security of the cryptoscheme used by \sys follows from these results. 
\begin{proposition}
\label{lemma1}
Assume that \sys uses \textsc{ckks} encryptions with parameters $(\mathcal{N},\ \Delta,\ \eta,\ mc)$ ensuring post-quantum security. Given a passive adversary corrupting at most $s-1$ parties out of $s$ parties in total, \sys achieves \emph{input confidentiality}.
\end{proposition}

\descr{Sketch of the Proof.} We consider a real-world simulator $\mathcal{S}$ that simulates the view of a computationally-bounded adversary corrupting $s-1$ parties, i.e., it has access to the inputs and outputs of $s-1$ parties. In Step 1 of \sys's workflow (Alg.~\ref{alg:workflow}), the simulator obtains the public parameters and the entire matrix $\bm{\tilde{A}}$, except the rows that belong to the honest DP. From Step 1 to the end, the DPs exchange only collectively encrypted information. In Step 8, each DP projects its local data on the obtained collectively encrypted PCs. If required by the application, the collectively encrypted result is switched to each DP's public key so that they can decrypt the final result. To avoid information leakage about data and/or about the encryption keys from the processed ciphertexts \cite{li2021security}, we rely on existing countermeasures \cite{mouchet2019distributedbfv,cheon2020remark,de2020fast} and add fresh noise (i.e., re-randomizing) sampled from a distribution that has a variance significantly larger than that of the input ciphertext's noise distribution to the processed ciphertext \cite{li2021security}. Alternatively, this result can be kept encrypted and used for future steps without ever being decrypted. Hence, by generating random ciphertexts with parameters $(\mathcal{N},\ \Delta,\ \eta,\ mc)$, $\mathcal{S}$ can simulate all the values communicated during the entire process such that the real outputs cannot be distinguished from the ideal ones. The sequential composition of all cryptographic functions remains simulatable by $\mathcal{S}$ as there is no dependency between the random values that an adversary can exploit. Also, the adversary cannot decrypt collectively encrypted data unless all DPs collude, which would contradict the considered threat model (\S.\ref{sec:overview}). Following this, \sys ensures \textit{input confidentiality} for the honest DP(s).

\section{\df{Extensions}}\label{app:extensions}

\renewcommand{\thesectiondis}[0]{\Alph{section}.}
\vspace{-0.5em}
\subsection{\df{Hybrid Use of Multiparty Security Primitives}}
\label{app:mheLss}
\vspace{-0.5em}
 \sys's multiparty construction enables it to seamlessly switch between MHE and secure multiparty computation (SMC) primitives based on secret sharing \cite{keller2018overdrive, bogdanov2008sharemind, shamir1979share}. Intuitively, an MHE ciphertext is transformed to linear (additive) secret shares (LSS) through a collective masked decryption by the DPs, i.e., each DP partially decrypts the ciphertext and masks the result with its secret share, whereas the last DP decrypts and obtains its share. After the computations in the LSS domain, to transform the result back to an MHE ciphertext, each DP encrypts its local share of the result such that it can be aggregated under MHE with all DPs' encrypted shares. We detail these procedures in Protocol~1. \sys always employs MHE to execute large-dimensional matrix operations and can perform non-polynomial operations, e.g., square root and divisions (in Alg.~\ref{alg:hh}), as well as small-matrix operations (in Alg.~\ref{alg:eigendecompo}), using LSS-based routines \cite{cho2018secure}. This combines the strengths of both approaches: On the one hand, relying on edge-computing and the SIMD property of MHE, \sys efficiently performs vectorized and parallel operations over large encrypted matrices while minimizing communication. On the other hand, relying on LSS-based SMC, \sys simplifies its usage by removing the need to choose intervals for non-polynomial function approximations. Note that efficient protocols exist for computing the bit-length of a secret-shared value \cite{dahl2012secure}, which can be used to map the input to a common interval for accurate approximation. In addition, LSS-based routines can be more efficient for computation over small data, e.g., eigendecomposition of a tiny matrix in RPCA, where the ciphertext packing is underutilized for MHE. In \sys, all costly non-polynomial operations (e.g., see Alg.~\ref{alg:hh}) are executed on a single scalar input. Thus, executing these operations on compact secret-shared data can further reduce the computational cost of \sys. 

\descr{LSS Scheme.}  We implemented a collection of SMC routines used by the prior work on LSS-based PCA~\cite{cho2018secure} to perform the required non-linear operations (comparison, square root, and division), newly extending the support to more than two DPs and 128-bit security.
 These protocols build upon a combination of existing SMC techniques~\cite{bogdanov2008sharemind, beaver1991efficient, shamir1979share, dahl2012secure, catrina2010secure,markstein2004software,nishide2007multiparty}. All values and shares are encoded as field elements $\bar{x}$ (or $\bar{\bm{x}}$ for a vector of elements) in $\mathbb{Z}_{\bar{p}}$, with $\bar{p}$ a prime, by relying on a fixed-point representation \cite{catrina2010secure}. For example, with 2 parties, $x \in \mathbb{Z}_{\bar{p}}$ is secret shared as $r  \in \mathbb{Z}_{\bar{p}}$ and $(x-r) \in \mathbb{Z}_{\bar{p}}$. Additions consist in simple share additions, whereas multiplications are done by relying on Beaver multiplication triples \cite{beaver1991efficient}. Following the prior work, we adopt the server-aided model of preprocessing whereby a third-party generates these triples to facilitate the main interactive computation with efficiency. This scheme can be modified to avoid the need of a trusted node at setup, thus following \sys's default threat model, by relying on an interactive protocol for the setup to be executed among all DPs. Adapting existing solutions \cite{keller2018overdrive, mouchet2019distributedbfv} to \sys is part of future work.  

\descr{Protocol to Switch Between MHE and LSS.} We build on the collective bootstrapping protocol \cite{mouchet2019distributedbfv}. We split this protocol in two rounds and add a conversion to/from the field $\mathbb{Z}_{\bar{p}}$ of the LSS scheme, see Protocol~1. We assume that DP$_1$ wants to perform a function $f_{LSS}$ on the encrypted vector $\bm{c}$. 
The security of the protocol can be derived from the security of the original $\text{DBootstrap}$ (for which Froelicher et al. \cite{spindle} prove that statistical indistinguishability is preserved as long as the masks are sampled from the correct distribution), from the LSS scheme guarantees (i.e., statistical indistinguishability), and from the security of \textsc{ckks} and the hardness of the decisional-RLWE problem~\cite{cheon2017homomorphic,lyubashevsky2010ideal, Lindner2011}.

\descr{Evaluation.} Switching to LSS removes the need for defining approximation intervals to evaluate non-polynomial functions hence simplifies the usage of \sys. Depending on the setting, it can also improve \sys's accuracy. The intervals for the non-polynomial operations depend on the DPs' data and, as \sys's intermediate results are repeatedly orthogonalized (through $\text{QR}^T$), these ranges can be accurately inferred upfront by the DPs, e.g., by simulating the protocol (\S.\ref{subsubsec:approx}).
\begin{protocol}{MHE $\Longleftrightarrow$ LSS}
\footnotesize
    \setcounter{protocol}{\value{protocol}-1}
	\refstepcounter{protocol}
	\renewcommand{\thealgorithm}{}
	\vspace{-1.0em}
	\footnotesize
	\textbf{Input:} DP$_1$ has $\bm{c}_{pk} = (\bm{c}_{0}, \bm{c}_{1}) = \{\bm{c}, \tau, \Delta\} \in R_{Q_{\tau}}^{2}$ a ciphertext encrypting $\bm{\tilde{p}}$. $\nu$ is a security parameter, $sk_{i}$ the secret-key of each $DP_i$, $\chi_{err}$ a distribution over $R$, where each coefficient is independently sampled from Gaussian distribution with the standard deviation $\sigma = 3.2$, and bound $\lfloor{6\sigma}\rfloor$. $\text{Encode}(\cdot)$ is the mapping from a plaintext encoded in $R$ to the equivalent encoding in $\mathbb{Z}_{\bar{p}}$. Let $T$ be the bound on all possible coefficients in the polynomial representation of $\bm{\tilde{p}}$ encoding real data values and $l$ be the bound on the possible bit length of real data values encoded in $\mathbb{Z}_{\bar{p}}$.\\
	\footnotesize
	\textbf{Output:} $\bm{c}'_{pk} = \{\bm{c}', L, \Delta\}$\\
    \footnotesize
    \textbf{Constraints:} $Q_{\tau}>(s+1)\cdot T \cdot 2^{\nu}$ \& $(s+1) \cdot 2^{\nu + l}$ < $\bar{p}$\\
	\begin{algorithmic}[1]
	\footnotesize
	\vspace{-1.0em}
	\STATE DP$_1$ broadcasts $\{\bm{c}, \tau, \Delta\}$
	\STATE Each DP$_i$ for $i = 2,\ \dots,\ s$:
	\STATE \quad Samples $\bm{a}_{i} \leftarrow \text{Uniform}(R_{T \cdot 2^{\nu}})$, $e_{i} \leftarrow \chi_{err}$
	\STATE \quad Sends $\bm{h}_{i} = sk_{i} \cdot \bm{c}_{1}+\bm{a}_{i}+e_{i} \ mod(Q_{\tau})$ to DP$_1$
	\STATE \quad Assigns $\bar{\bm{a}}_{i} = \text{Encode}($- $\bm{a}_{i})\ mod(\bar{p})$
	\STATE DP$_1$:
	\STATE \quad Samples $\bm{a}_{1} \leftarrow \text{Uniform}(R_{T \cdot 2^{\nu}})$, $e_{1} \leftarrow \chi_{err}$
	\STATE \quad Computes $\bm{h}_{1} = sk_{1} \cdot \bm{c}_{1}+\bm{a}_{1}+e_{1} \ mod(Q_{\tau})$
	\STATE \quad Computes $\bm{h'} = \bm{c}_{0} + \sum_{i=1}^{s} \bm{h}_{i} \ mod(Q_{\tau})$
	\STATE \quad Assigns $\bar{\bm{a}}_1$ = $\text{Encode}(\bm{h'} - \bm{a}_1)\ mod(\bar{p})$
	\STATE All DP$_i$: 
	\STATE \quad Compute $\bar{\bm{r}}_i = f_{LSS}(\bar{\bm{a}}_i)$
	\STATE \quad Compute $\bar{\bm{r}}_i = \bar{\bm{r}}_i - \bar{\bm{b}}_i\ mod(\bar{p})$, with $\bar{\bm{b}}_i \leftarrow \text{Uniform}(\mathbb{Z}^{\mathcal{N}/2}_{2^{\nu+l}})$ 
	\STATE \quad Encrypt $\bm{c}^{(i)}_{pk} =$ Enc($pk, \bar{\bm{b}}_i$)
	\STATE Each DP$_i$ for $i = 2,\ \dots,\ s$: Sends $\bm{c}^{(i)}_{pk}$ and $\bar{\bm{r}}_i$ to DP$_1$
	\STATE DP$_1$:
	\STATE \quad Computes $\bar{\bm{r}}' = \sum_{i=1}^{s} \bar{\bm{r}}_i \ mod(\bar{p})$ and encrypts $\bm{c}'_{pk} = $ Enc($pk, \bar{\bm{r}}'$)
	\STATE \quad Computes $\bm{c}'_{pk} =  \bm{c}'_{pk} +  \sum_{i=1}^{s} \bm{c}^{(i)}_{pk}$
	\vspace{-1.5em}
	\end{algorithmic}
	\label{prot:MHESES}
\end{protocol}
For example, with the MNIST dataset and the parameters of Tab.~\ref{tab:accuracy}, the DPs define 16 distinct intervals to evaluate 1,047 polynomial approximations. This is because the ranges of values are constant across the dimensions and across the iterations of the same operations. For the same dataset, relying on \sys\textsc{+lss} improves the Pearson correlation between \sys's PCs and the PCs obtained with a standard non-secure centralized PCA from 0.91 to 0.92 (when using our default parameters, Tab.~\ref{tab:symbols}). \sys's accuracy depends on the size and degree of the intervals hence can be improved by refining these parameters. We illustrate this on the execution of a single $\text{QR}^T$ in Tab.~\ref{tab:localruntimesSES}. We note that QR$^T$ is used (iteratively) in Steps 4, 6 and 7 of \sys and that all non-polynomial operations in \sys are executed in the Householder (HH, Alg.~\ref{alg:hh}) that is called in line 2 of QR$^T$. For QR$^T$ on a 8x8 matrix, HH is called seven times and requires the evaluation of three non-polynomial functions. In \sys, this requires the definition of 21 approximation intervals, i.e., one per non-polynomial function. We show that using a single large interval (with a polynomial of degree 63; [0:1000;63]) for all operations already yields results that are correlated with the results obtained by a cleartext solution. \sys's accuracy can then be improved by either downsizing the interval (to [0:100;63]), increasing the approximation degree (to [0:1000;127]), or by using more fine-grained intervals for the different steps in the computation ([0:100;63] for the first execution of HH and [0:1;63] afterwards).

In Fig.~\ref{fig:performanceDPsSES}, we show that \sys's runtime is similar with or without this extension. \sys's computational cost is reduced by computing on secret shares, instead of on encrypted vectors, but this gain is overshadowed by the communication overhead brought by both the protocol for switching between MHE and LSS and by LSS distributed computations. \sys scales similarly with its default approach (\sys in Fig.~\ref{fig:performanceDPsSES}) and when switching to LSS for non-polynomial and small-dimensional operations (\sys\textsc{+lss}). Switching to LSS only for non-polynomial operations (\sys\textsc{+lss-op}) is around 1.4x slower than \sys\textsc{+lss} due to the communication overhead brought by the high-number of switches between the two schemes. \sys can optimize its runtime for the small-dimensional eigendecomposition (Step 6) by performing it entirely in the LSS domain, which is up to 1.5x faster than in its basic approach in this scenario. When operating on larger dimensions, i.e., in Step 4 (Alg.~\ref{alg:workflow}), \sys only switches to LSS for small-dimensional (i.e., single value as shown in Alg.~\ref{alg:hh}) non-polynomial operations as this can improve its precision. Performing sequences of operations in LSS in Step 4 would require to switch and operate on large-dimensional secret-shared elements, which would further increase the communication overhead. In Tab.~\ref{tab:localruntimesSES}, we show that the runtimes of most LSS operations are in the same order of magnitude as MHE operations.

\begin{figure*}[ht!]
\vspace{-3.0em}
        \scriptsize
        \center
        \subfloat[\footnotesize \textbf{Runtime with LSS.}]{
        \includegraphics[width=0.20\textwidth,valign=b]{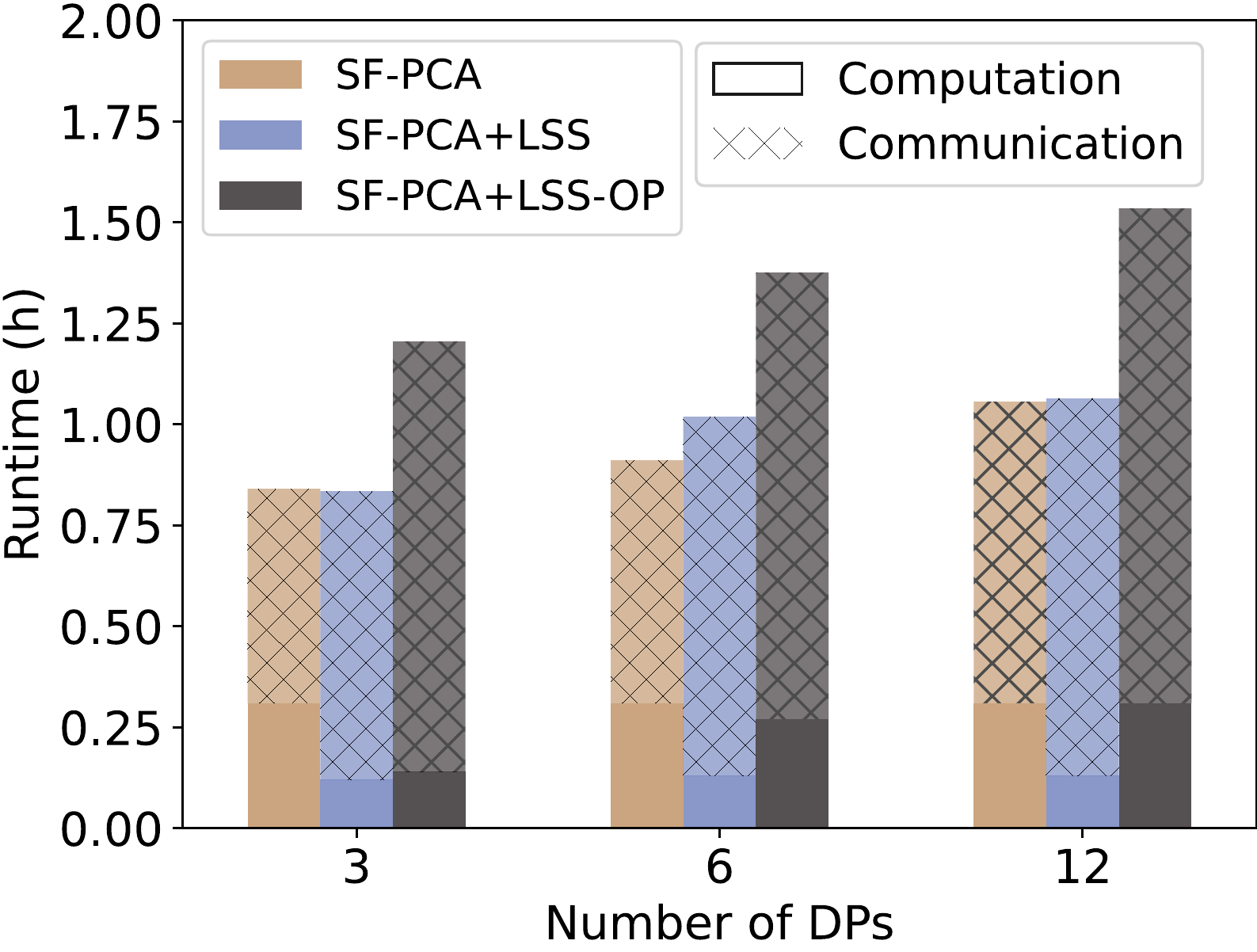} \vspace{-1.0em}
        \label{fig:performanceDPsSES}}
        \subfloat[ \footnotesize \textbf{Accuracy (MSE;$r^2$) and runtimes (seconds) for MHE and LSS operations.}]{
         \adjustbox{valign=b}{\setlength\tabcolsep{1.5pt}\begin{tabular}{ |c|c|c|c|c| }
            \hline
            \textbf{Operation} & \textbf{HE-Runtime} & \textbf{HE-Accuracy} & \textbf{LSS-Runtime} & \textbf{LSS-Accuracy}  \\ \toprule
            \hline
            M5 & $0.9$ & - & $0.09$ & - \\ \hline
            QR$^T(\bm{M}^{(8 \times 8)})$ w. [0:1000; 63] & $117$ & $10^{-2};0.86$ & $90$ & $10^{-8};0.99$\\ \hline
            QR$^T(\bm{M}^{(8 \times 8)})$ w. [0:1000; 127] & $117$ & $10^{-3};0.99$ & $90$ & $10^{-8};0.99$\\\hline
            QR$^T(\bm{M}^{(8 \times 8)})$ w. [0:100; 63] & $117$ & $10^{-4};0.99$ & $90$ & $10^{-8};0.99$\\ \hline
             QR$^T(\bm{M}^{(8 \times 8)})$ w. [0:100, 0:1; 63] & $117$ & $10^{-5};0.99$ & $90$ & $10^{-8};0.99$\\\hline
            \textsl{Send}($c$ or share) & $0.026$ & - & $0.021$ & - \\ \hline
            Switch MHE $\leftrightarrow$ LSS & $0.762$ & - & - & - \\ \hline
            \end{tabular}}\label{tab:localruntimesSES}\vspace{-0.5em}}
            \subfloat[\footnotesize \textbf{Runtime w. uneven data split.}]{
        \includegraphics[width=0.20\textwidth,valign=b]{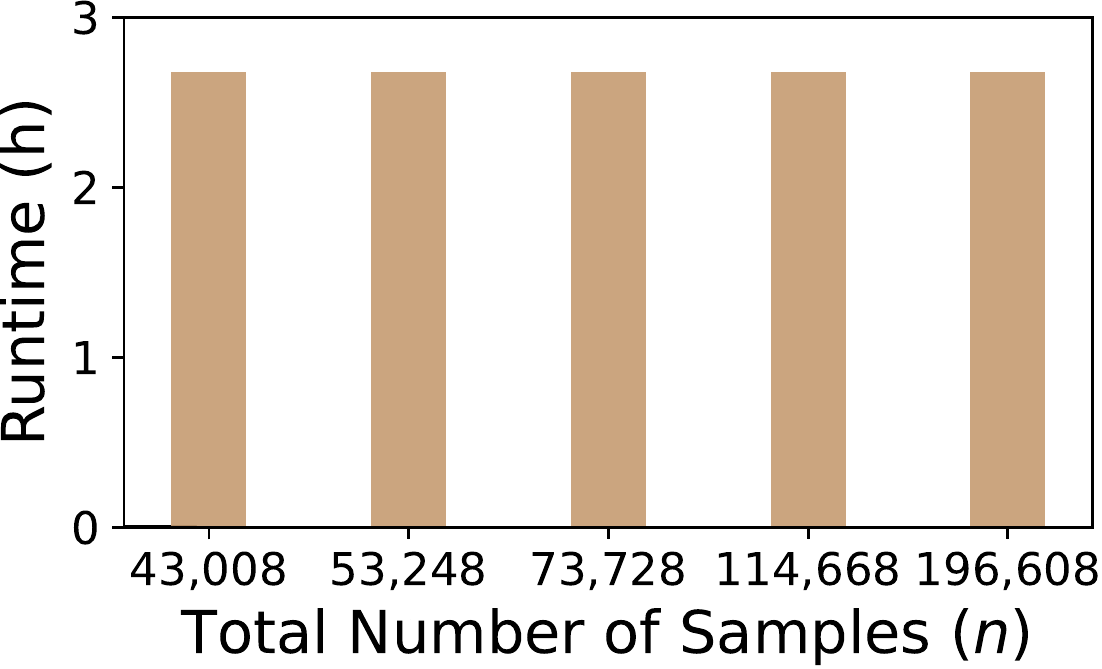} \vspace{-1.0em}
        \label{fig:maxNbrOfSamples}}
            \vspace{-0.5em}
        \caption{\footnotesize In Fig.~\ref{fig:performanceDPsSES}, we show \sys's runtime when using LSS extension. In Tab.~\ref{tab:localruntimesSES}, $[0:100; 63]$ indicates that the approximations are done in an interval between 0 and 100 with a degree of 63. Fig.~\ref{fig:maxNbrOfSamples} depicts \sys's runtime when one DP has 32,786 data samples and the remaining samples are evenly split among 5 DPs.}
        \vspace{-1.5em}
\end{figure*}
\vspace{-0.5em}
\subsection{\df{Vertically Partitioned Data}}\label{app:verti}
\vspace{-0.5em}
\sys's workflow can be easily adapted to work with a vertically partitioned input matrix by modifying the interactive computation among the DPs while leveraging the same local operations as before. Because of the different way the data is split, some of the DPs' intermediate results have to be combined (aggregated or concatenated) at different points in \sys's workflow; this does not change the nature of the underlying operations that are optimized in the default setting of \sys. In the vertical case, the overall computation and communication complexities depend on the total number of samples $n$ and the number of features per DP $m_i$, whereas these depend on $n_i$ and $m$ in \sys's original approach. We show both approaches in Fig.~\ref{fig:randoPCA_Verti}. The mean-centralization is up to eight times less expensive than in \sys's original workflow, because each DP$_i$ keeps its part of the averages' vector $\bm{\tilde{o}}_i$ in cleartext.

\begin{figure}[ht!]
\vspace{-0.5em}
    \centering
    \small
    \includegraphics[width=1\columnwidth]{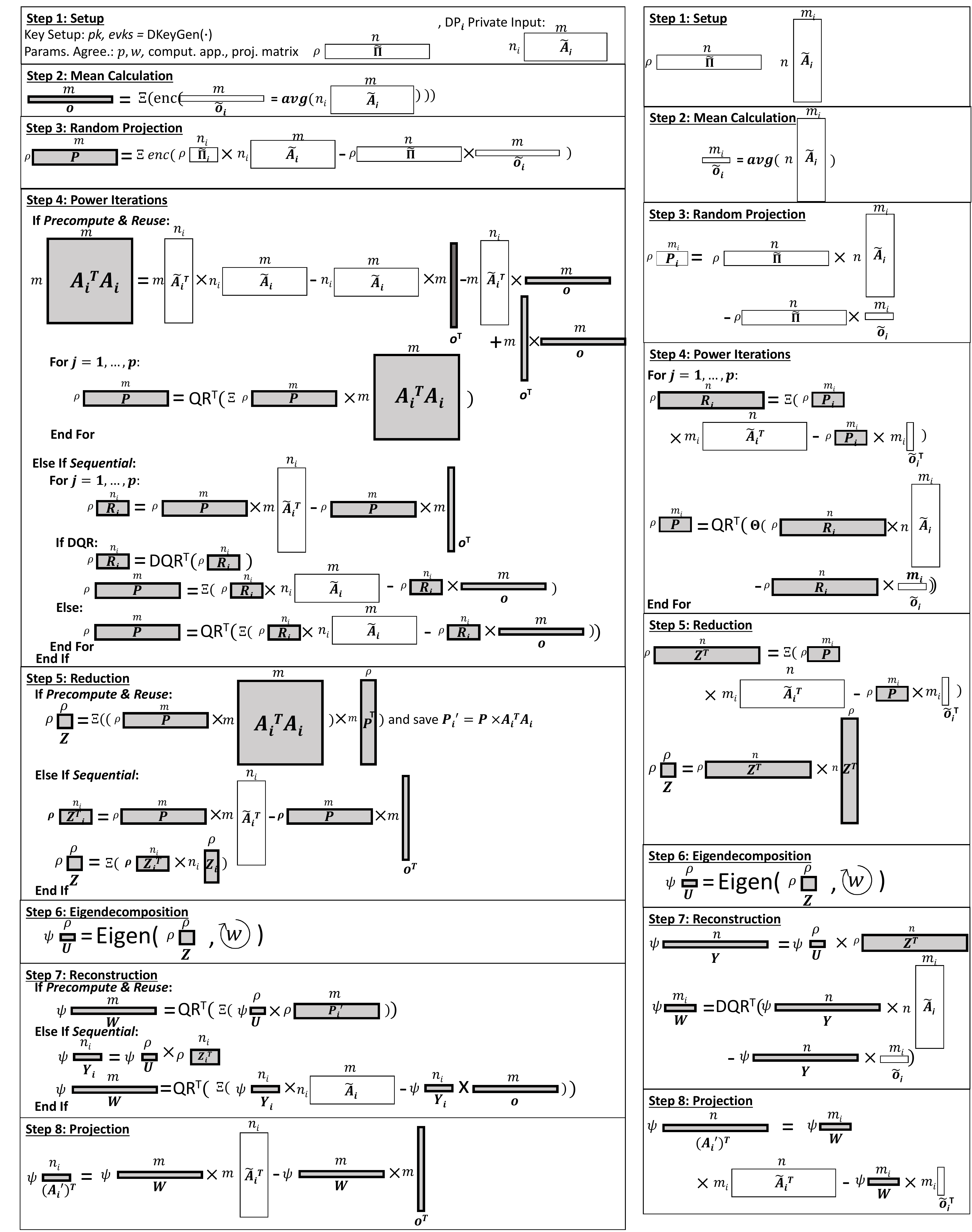}
    \vspace{-1.5em}
    \caption{\footnotesize\textbf{\sys's secure workflow with horizontally (left) and vertically (right) partitioned data.} The execution is depicted from DP$_i$'s point of view. The filled boxes indicate encrypted matrices and the empty boxes show cleartext matrices. The dimensions are shown with the box sizes and are indicated on the left and top of the corresponding box. $\Xi$ indicates a collective aggregation, $\Theta$ a collective concatenation, and, in both cases, the result is broadcast to all DPs. The dimensions equal to 1 are omitted and vectors replicated to comply with the matrix-multiplication dimensions are not shown. \df{A tilde indicates cleartext.}}
    \label{fig:randoPCA_Verti}
\end{figure}


\vspace{-0.5em}
\subsection{\df{Differential Privacy}}\label{app:diffP}
\vspace{-0.5em}
\df{\sys can be extended to provide differential privacy by leveraging an interactive protocol in which the DPs sequentially shuffle an encrypted list of noise values before adding them to the results upon decryption \cite{drynx}. The choice of privacy parameters and maintaining accuracy are part of future work.} 

\vspace{-0.5em}
\subsection{\df{Fault Tolerance}}\label{app:fault}
\vspace{-0.5em}
To cope with the possibility of a subset of DPs becoming unavailable during the PCA computation, \df{which is particularly relevant when there are many DPs,} \sys can be extended by employing a $\hat{s}$-out-of-$s$ threshold secret-sharing for the MHE secret keys \df{\cite{mouchet2022efficient}}, where $s$ is the number of DPs. Note that the main setting of \sys considers $\hat{s}$\,=\,$s$. Setting $\hat{s}$ to be smaller than $s$ changes \sys's security model to tolerate up to $\hat{s}-1$ dishonest DPs. As long as at least $\hat{s}$ DPs are available for each interactive step, \sys's execution continues without interruption. In certain steps of \sys the omission of a subset of parties may result in their local data not being accounted for in the computation. However, given the iterative nature of the RPCA algorithm, the overall results are expected to be robust against such omissions with a sufficient number of iterative steps.

\renewcommand{\thesectiondis}[2]{\Alph{section}:}
\vspace{-0.5em}
\section{Datasets} \label{app:datasets}
\vspace{-0.5em}
The Wine dataset \cite{wine} contains 4,898 wine samples with physicochemical attributes as features and a quality score as label. The Lung dataset \cite{lungcancer} contains 9,098 patients with 23,724 genomic variations (as features) and a label indicating the presence of a cancer. The PIMA dataset ($768 \times 8$)~\cite{pima} contains medical observations collected from an Indian community that can be used to predict the presence of diabetes. Chr20 ($2,502 \times 1,773$)~\cite{1000genomes} is a subset of the genomic data available in the 1,000 Genomes dataset. In the MNIST dataset ($70,000 \times 784$)~\cite{MNIST}, each sample describes the grey-scale image of a single handwritten digit. Finally, the Vehicle \cite{vehicle} dataset contains 19 features extracted from each of the 435 images of buses or cars.

\vspace{-0.5em}
\section{\df{Runtime Scales with the Slowest DP}}\label{app:maxNbrOfSample}
\vspace{-0.5em}

\df{We show in Figure \ref{fig:maxNbrOfSamples} that \sys's runtime depends on the maximum number of local samples among the DPs. In this example, the DP with the maximum number of samples has 32,768 samples and the other samples are evenly split among the remaining 5 DPs. Even as the total number of data samples increases, \sys's runtime remains constant since the maximum number of local samples stays the same.}

\section{Using \sys to Improve Machine Learning Efficiency and Accuracy} \label{app:secureML}
By combining \sys with a privacy-preserving solution for a downstream machine learning (ML) task, a secure federated ML workflow supporting the full analytic pipeline, encompassing pre-processing (e.g., dimension reduction), training, and inference, can be built.
For example, \sys can be seamlessly integrated with existing MHE-based solutions for training generalized linear models~\cite{spindle,zheng2019helen} or neural networks (NNs)~\cite{sav2020poseidon}. As the training time of these solutions increase with the number of features in the dataset, \sys may be a useful solution for reducing the scale of high-dimensional datasets to speedup model training. For example, executing \sys on the MNIST dataset (see Tab.~\ref{tab:accuracy}) to project it on 5 PCs takes 1 hour and reduces the number of features by a factor of 152. Training a model using the PCs instead of the original features would reduce by a factor 7 the runtimes of previously mentioned secure solutions. Such an approach can also improve the accuracy of ML models when dimension reduction results in noise removal and more informative features, especially in limited data settings~\cite{accuracy1, accuracy2, accuracy3, accuracy4}.
We illustrate this use case by training a NN model (i.e., a multilayer perceptron with hidden layer made of four nodes, sigmoid activation functions, and one output node) to perform classification on the Vehicle dataset \cite{vehicle}, which contains 435 samples with 18 features derived from vehicle images (Appendix \ref{app:datasets}). We observed that the model trained without any preprocessing achieves a prediction accuracy of $55\%$ on the test set, whereas training the model on 5 PCs obtained by \sys (applied to the training data) as features yields an accuracy of $87\%$, which increases to $97\%$ with 8 PCs. This small example illustrates the fact that by de-correlating the features and reducing their number, PCA can improve ML model accuracy.

\end{document}